\documentclass[twocolumn]{aastex63}
\usepackage{apjfonts}
\usepackage{amsmath}
\usepackage{multirow}
\usepackage{graphicx}
\usepackage{threeparttable}
\usepackage{hyperref}
\usepackage{lineno}
\usepackage{soul}

\shorttitle{N/S Asymmetry and the Phase Space Snail}
\shortauthors{Guo et al.}

\begin{document}

\title{The North/South Asymmetry of the Galaxy: Possible Connection to the Vertical Phase Space Snail}

\author{Rui Guo}

\affiliation{Department of Astronomy, School of Physics and Astronomy, Shanghai Jiao Tong University, 800 Dongchuan Road, Shanghai 200240, China}
\affiliation{Key Laboratory for Particle Astrophysics and Cosmology (MOE) / Shanghai Key Laboratory for Particle Physics and Cosmology, Shanghai 200240, China}

\author{Juntai Shen}
\affiliation{Department of Astronomy, School of Physics and Astronomy, Shanghai Jiao Tong University, 800 Dongchuan Road, Shanghai 200240, China}
\affiliation{Key Laboratory for Particle Astrophysics and Cosmology (MOE) / Shanghai Key Laboratory for Particle Physics and Cosmology, Shanghai 200240, China}

\author{Zhao-Yu Li}
\affiliation{Department of Astronomy, School of Physics and Astronomy, Shanghai Jiao Tong University, 800 Dongchuan Road, Shanghai 200240, China}
\affiliation{Key Laboratory for Particle Astrophysics and Cosmology (MOE) / Shanghai Key Laboratory for Particle Physics and Cosmology, Shanghai 200240, China}

\author{Chao Liu}
\affiliation{Key Lab of Space Astronomy and Technology, National Astronomical Observatories, Chinese Academy of Sciences, 20A Datun Road, Chaoyang District, Beijing 100101, China}

\author{Shude Mao}
\affiliation{Department of Astronomy, Tsinghua University, Beijing 100084, China}
\affiliation{National Astronomical Observatories, Chinese Academy of Sciences, 20A Datun Road, Chaoyang District, Beijing 100101, China}

\correspondingauthor{Juntai Shen, Zhao-Yu Li}
\email{jtshen@sjtu.edu.cn; lizy.astro@sjtu.edu.cn}

%



\begin{abstract}
The Galaxy is found to be in disequilibrium based on recent findings of the North/South (N/S) asymmetry and the phase mixing signatures, such as a phase spiral (snail) structure in the vertical phase space ($z-V_{z}$). We show that the N/S asymmetry in a tracer population of dwarfs may be quantitatively modeled with a simple phase snail model superimposed on a smooth equilibrium background. As the phase snail intersects with the $z$ axis, the number density is enhanced, and the velocity dispersion ($\sigma_{z}$) is decreased relative to the other side of the Galactic plane. Fitting only to the observed asymmetric N/S $\sigma_{z}$ profiles, we obtain reasonable parameters for the phase space snail and the potential utilized in modeling the background, despite the complex dependence of the model on the potential parameters and the significant selection effects of the data. Both the snail shape and the N/S number density difference given by our best-fit model are consistent with previous observations. The equilibrium background implies a local dark matter density of $0.0151^{+0.0050}_{-0.0051}$ ${\rm M}_{\odot}\,{\rm pc}^{-3}$. The vertical bulk motion of our model is similar to the observation, but with a $\sim$1.2 $\rm km\,s^{-1}$ shift. Our work demonstrates the strong correlation between the phase space snail and the N/S asymmetry. Future observational constraints will facilitate more comprehensive snail models to unravel the Milky Way potential and the perturbation history encoded in the snail feature.
\end{abstract}



\section{Introduction} 
\label{sec:intro}
Since the pioneering work of \cite{Oort1932}, many studies have tried to infer the local surface density and vertical force by analyzing the stellar kinematics in the Solar Neighborhood \citep[e.g.][]{Bahcall1984, Kuijken1989a, Kuijken1989b, Holmberg2000, Bovy2012}. Combined with measurements of the rotation curve and the local baryon density, the vertical force can be utilized to estimate the local dark matter density \citep[see reviews in][]{Read2014, Salas2020}. In the studies based on the local stellar kinematics, the distribution function and the vertical Jeans equation have been widely used to derive the local dark matter density \citep[e.g.][]{Kuijken1989a, Kuijken1989b, Holmberg2000, Holmberg2004, Garbari2012, Zhang2013, Bienayme2014, Xia2016, Sivertsson2018, Widmark2019, Guo2020}. These studies have a common and basic assumption, i.e. the local Galactic disc is in dynamical equilibrium. However, non-equilibrium phenomena of the vertical structure in the Solar Neighborhood have been observed as a wave-like perturbation in the vertical stellar counts and a vertical bulk motion \citep[e.g.][]{Widrow2012, Williams2013, Carrillo2018}, and more recently as a snail feature in the vertical phase-space distribution \citep[e.g][]{Antoja2018, Bland2019, Laporte2019, LiZY2020}.

The first evidence of the local disequilibrium in the vertical dynamics is the North-South asymmetry (hereafter N/S asymmetry) in both the spatial density and vertical velocity discovered by \cite{Widrow2012}. The number density contrast between the North and South, i.e. (North-South)/(North+South), shows a wave-like pattern, which reveals a number of deficits and peaks at some vertical heights. This number count asymmetry was confirmed by \cite{Yanny2013} with a careful analysis of the uncertainties, and by \cite{Bennett2019} with a larger sample size and more accurate distance measurements. It is also found that the asymmetry is independent of the stellar color, which implies that it is intrinsic to the disc.

The spatially dependent bulk motions in the vertical direction were also detected by several groups using different surveys \citep{Widrow2012, Williams2013, Carlin2013, Carrillo2018, Bennett2019}. These vertical perturbations are classified into bending and breathing modes according to the parity in density and vertical velocity \citep{Widrow2014}. The breathing modes can be caused by non-axisymmetric features in the disk, such as the spiral arms or the Galactic bar \citep{Debattista2014, Faure2014, Monari2015, Monari2016}, while the bending modes are likely caused by external perturbations \citep{Gomez2013, Widrow2014, Chequers2018, Laporte2018}. Breathing modes can also be generated by external perturbation with fast fly-by interactions \citep{Widrow2014}. The vertical perturbation modes can help us understand the perturbation history of the Milky Way disc. However, the perturbation modes of the Galactic disc are still under debate. Earlier studies found a rarefaction-compression behaviour from inner Solar radii to outer radii \citep{Williams2013, Carlin2013}. More recently, \cite{Carrillo2018} found that the structure in the bulk motion pattern of vertical velocity ($V_{z}$) depends strongly on the adopted proper motions, and can be artificially created due to distance uncertainties. They suggested a combination of bending and breathing modes that may be of several different origins. This is also supported by the recent Gaia Data Release 2 (DR2) disk kinematics measurements \citep{Gaia2018a}. The bulk motion amplitudes found in Gaia DR2 are only less than 3 $\rm km\,s^{-1}$ in the vertical range of $|z| <1$ kpc.

A more striking indication of the vertical disequilibrium is the snail (spiral) features in the $z-V_{z}$ phase space discovered by \cite{Antoja2018} with the unprecedented Gaia DR2 \citep{Gaia2018b}. The phase space snail is interpreted as incomplete phase mixing in the Galactic vertical direction. It shows up more clearly when color-coded with the azimuthal velocity ($V_{\phi}$) and in the number density contrast map \citep{Laporte2019}. This indicates a strong correlation between the in-plane and vertical motions. The variations of the phase space snail with stellar age, chemistry, and orbital properties have been extensively investigated by many following works \citep[e.g.][]{Tian2018, Bland2019, ChengXL2019, Laporte2019, LiZY2020}. The shape of the snail also varies radially and azimuthally \citep[e.g.][]{Bland2019, Laporte2019, WangC2019, XuY2020, LiZY2021}. The snail is found to wrap up more tightly in smaller Galactic radii, and be more elongated along the $z-$axis in larger radii \citep{Bland2019, XuY2020, LiZY2021}. \cite{LiZY2020} found that the snail shell in the Solar Neighborhood exists in colder orbits only; the hotter orbits show a rather smooth $z-V_{z}$ phase space distribution, suggesting that the phase mixing in hotter orbits probably has finished. Besides, the clarity of the snail shape is found to increase when the stars are grouped by their guiding radii \citep{LiZY2021}.

Several works have attempted to explain the phase space snail by the passage of a satellite galaxy, such as the last pericentric passage of the Sagittarius dwarf \citep{Antoja2018, Binney2018, Bland2019, Laporte2019, Bland2021}. Note that, the interaction of  satellite galaxies or subhalos with the Milky Way can also perturb the disc to generate the vertical bulk motions, i.e. the  bending and breathing modes in the outer disc \citep{Purcell2011, Gomez2013, Widrow2014, Laporte2018, Bland2021}. In \cite{Darling2019}, pure vertical bending waves in a stellar disc can evolve to form phase space snails similar to the observations. Other internal perturbations, such as the buckling instability of the bar, are also possible to produce phase space snails \citep{Khoperskov2019}. However, the bar-driven phase snail scenario is not supported by \cite{LiZY2020} based on the absence of phase snail on the radial velocity ($V_{R}$) color-coded phase space. Studying the snail shapes at different Galactic locations can help to distinguish different origins. It can also help to constrain the perturbation time, such as by analyzing the residuals in the frequency-angle space shown in \cite{LiH2021}. In addition, based on the absence of the phase snail in hotter orbits, \cite{LiZY2020} constrained that the previous perturbation should be at least 500 Myr ago, or the hotter orbits will not have enough time to complete the phase mixing. More recently, \cite{LiZY2021} used the pitch angle of the phase snail to constrain the perturbation time to $\sim$500 - 700 Myr ago, consistent with previous results.

These observations of the asymmetries and phase space snail challenge the equilibrium assumption that is usually made in the vertical dynamical models. The influence of the departure from equilibrium to the determination of the total surface density and of the local dark matter density has been investigated in simulations by \cite{Banik2017} and \cite{Haines2019}. They argued that the disc perturbation will cause a systematic overestimate of both the vertical force and local density when the equilibrium is still assumed. In our previous work \citep{Guo2020}, we used a large sample of G-/K- dwarf stars to measure the local dark matter density with the vertical Jeans equation. We found that the asymmetric Northern and Southern velocity dispersion profiles will result in quite different estimates of the local dark matter density. In \cite{Salomon2020}, the local dark matter density in the less perturbed South is only slightly smaller than that value derived from the more perturbed North. The difference in the derived surface density between the North and South is small only at large vertical heights ($|z| > 2.2$ kpc). Closer to the disk plane, the surface density profiles still present a large discrepancy. Taken together, the nonequilibrium features need to be considered and new methods of deriving the local density need to be developed.

In order to recover the asymmetric Northern and Southern velocity dispersion profiles found in \cite{Guo2020} simultaneously, here we try to incorporate an analytical phase space snail in our dynamical model. We attempt to connect the asymmetries in the number density and vertical kinematics with the presence of the phase space snail, in addition to the underlying symmetric smooth distribution function. The idea naturally arises from the fact that when the phase snail intersects with $z$ axis at $z_{s}$, the number density at this vertical height will be enhanced, relative to the same vertical height ($-z_{s}$) on the opposite side of the Galactic plane. Meanwhile, when the snail intersects with $z$ axis, there are more stars from the snail concentrated near $V_{z} \sim 0$ $\rm km\,s^{-1}$, and thus the velocity dispersion will be relatively smaller. As a whole, the Northern and Southern number density profiles fluctuate with the opposite pattern, that is reversed in the velocity dispersion profiles. This is the phenomenon we observed in the previous work \citep{Guo2020}.

The paper is organized as follows: We present the dwarf sample selection and the test particle simulation in Section \ref{sec:data}. The models for the smooth background and the phase space snail are described in Section \ref{sec:models}. The results, including the fit to the asymmetric velocity dispersion profiles, the comparison of the snail shape, and the N/S number density differences derived from our snail models, are presented in Section \ref{sec:result}. Discussions and the conclusion are given in Sections \ref{sec:dis} and \ref{sec:con}, respectively.

\section{Data} 
\label{sec:data}
The coordinates system adopted in this paper is the classical Galactocentric cylindrical system ($R,\, \phi,\, z$), with the Galactocentric radius $R$ increasing radially outward, $\phi$ increasing in the direction of Galactic rotation, and positive $z$ pointing toward the North Galactic Pole. The Sun is located at ($-8.34, 0., 0.027$) kpc in the Galactic Cartesian coordinates system \citep{Reid2014b, ChenB1999}. The Solar motion relative to the local standard of rest (LSR) is $\rm (U_{\odot}, V_{\odot}, W_{\odot}) = (9.58, 10.52, 7.01)\ km\, s^{-1}$ \citep{TianH2015}, and the circular velocity of LSR is adopted as 240 km s$^{-1}$ \citep{Reid2014b}.

\subsection{Sample selection}
\label{ssec:dobs}
This work aims to recover the asymmetric vertical velocity dispersion profiles in our previous work \citep{Guo2020}, utilizing an analytical phase snail model. Thus the observational data we use here is still the G-/K- dwarf star sample selected from the cross-matched sample of LAMOST DR5 \citep{CuiX2012, ZhaoG2012} and Gaia DR2 \citep{Gaia2018b} (though Gaia eDR3 is available). This G-/K- type dwarf star sample consists of 93 609 stars, selected with criteria on the signal-to-noise ratio, the stellar surface gravity, the effective temperature, and the G-band magnitude etc. The quality constraints on the parallax $\varpi$ and parallax error $\sigma_{\varpi}$ are $\varpi > 0$ and $\sigma_{\varpi}/\varpi < 0.2$. Stars with $\rm [Fe/H] < -0.4$ are excluded to roughly exclude thick components. These criteria help to select a complete, single stellar population to be applied to the vertical Jeans equation. More details of the sample selection can be found in \cite{Guo2020}, and the spatial distribution of the sample in their Fig. 1.

The original spatial volume of this sample is constrained by $|z| < 1.3$ kpc, $|R - R_{\odot}| < 0.2 $ kpc and $|\phi| < 5^{\circ}$ (where $R_{\odot}$ is the Solar radius). In this work, we apply a slightly smaller vertical cut, i.e. $|z| < 1.0$ kpc, which will exclude $\sim$ 1400 stars. We have about two times larger sample size in the Galactic North than the South. The selection effects are corrected according to the Two Micron All Sky Survey (2MASS) \citep{Skrutskie2006}, with the distance error taken into consideration. The corrected number density is derived for each star, and then the vertical scale height is obtained by an exponential fitting to the number density profile. See Section 2.2 in \cite{Guo2020} for the detail process of selection effects correction.

When calculating the 3D velocities, the radial velocities $V_{\rm los}$ from LAMOST spectra are utilized. LAMOST radial velocities have a systematic offset $\sim 5.3$ ${\rm km\,s^{-1}}$ compared to Gaia radial velocities, which has been corrected in the calculation. When compared with the model velocity dispersion profile, the measurement error of 4.5 ${\rm km\,s^{-1}}$ \citep{GaoS2014} has been subtracted from the standard deviations of the vertical velocities.

\subsection{Snail phase space distribution inspired by a test particle simulation}
\label{ssec:tps}
Before fitting the observational data, we have also utilized a test particle simulation to build the snail models. This test particle simulation is taken from \cite{LiZY2021} using their first approach of imposing the vertical perturbation. The Milky Way potential of this simulation is taken as the Model I from \cite{Irrgang2013}. The test particles are sampled as a thin disc with a \emph{QuasiIsothermal} distribution function \citep{Binney2010, Binney2011}. The radial scale length and the vertical scale height of the disc are set as 3.7 kpc and 0.3 kpc, respectively. The vertical perturbation is imposed on all the test particles to mimic the external perturbation. This technique has been widely applied to study the evolution and radial variation of the phase space snail \citep[e.g.][]{Antoja2018, Binney2018, LiZY2020, LiZY2021}. The particles receive a vertical velocity kick simultaneously, while their vertical positions are barely changed. The random vertical velocity perturbation for each star has a Gaussian distribution with a median value of $-$30 km s$^{-1}$ and a dispersion of 5 km s$^{-1}$. Then, the test particles evolve and oscillate under the Milky Way potential to form phase space snails.

\begin{figure}[bth]
\centering
\includegraphics[width=0.8\columnwidth]{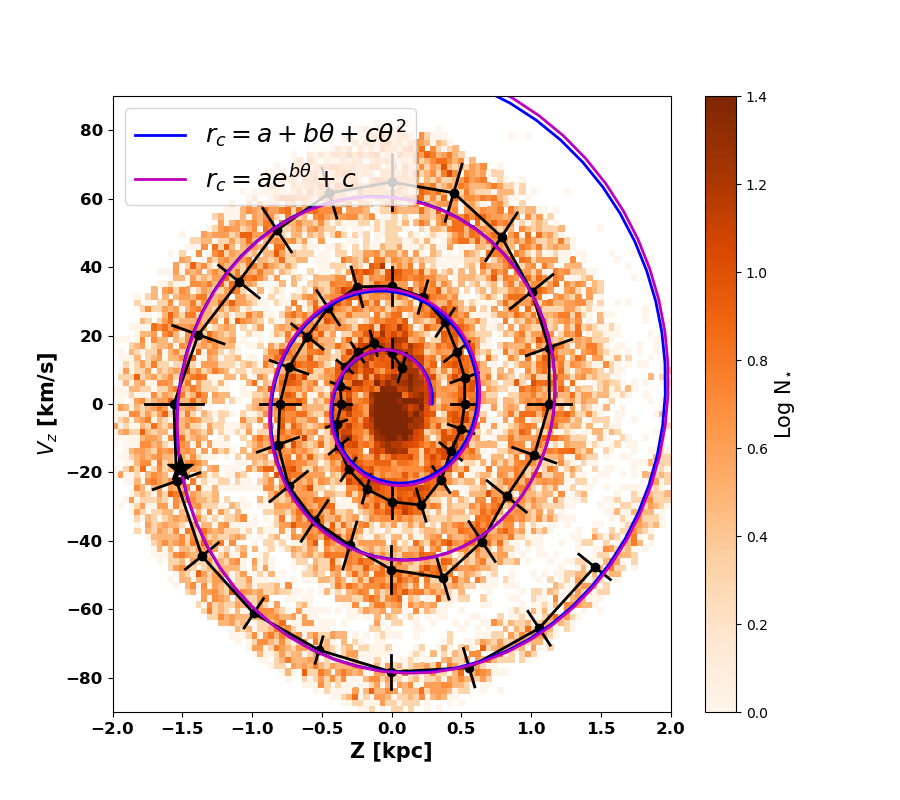}
\caption{
\label{fig:ms_2d}
Phase space distribution of the test particle simulation for a disk imposed with a vertical velocity kick. The radial range of particles shown here is the same as the observation, i.e. $|R-R_{\odot}| < 0.2$ kpc. The black points are the density peaks in different phase space wedges and snail wraps, while the error bars indicate the radial width of the snail. The blue and magenta lines are the fits to the black points with a quadratic and a logarithmic spiral function, respectively. The black star indicates the peak location of the angular distribution of the phase space snail.}
\end{figure}

Same as the observation, we have selected the test particles in a radial range of $|R - R_{\odot}| < 0.2 $ kpc at time t= 500 Myr. Their $z-V_{z}$ phase distribution is shown in Fig. \ref{fig:ms_2d}. It shows a clear phase snail feature extending to a vertical height of $\sim 2$ kpc and velocity of $\sim 90$ $\rm km\,s^{-1}$. In order to build an analytical model for this snail, we use the method in \cite{LiZY2021} to measure the phase snail shape (with a corresponding error bar to cover the width of the phase snail). It basically measures the density peaks in several wedges at different snail wraps, that are then connected to form the phase snail shape. We can then calculate the scaled phase space radii\footnote{Note that r stands for a dimensionless scaled phase space radius, which is different from the Galactocentric radius $R$.} as $r=\sqrt{(\frac{z}{1\, {\rm kpc}})^2 + (\frac{V_{z}}{45\, {\rm km/s}})^2}$ for the selected density peaks. Their phase space angles can be determined by manually adding $2k\pi$ ($k$ is an integer) according to their relative positions on the snail shell. It is more convenient to work with such analytical snail shape in $r-\theta$ space than directly fitting all the particles by simple analytical functions.

\begin{figure}[bth]
\centering
\includegraphics[width=0.9\columnwidth]{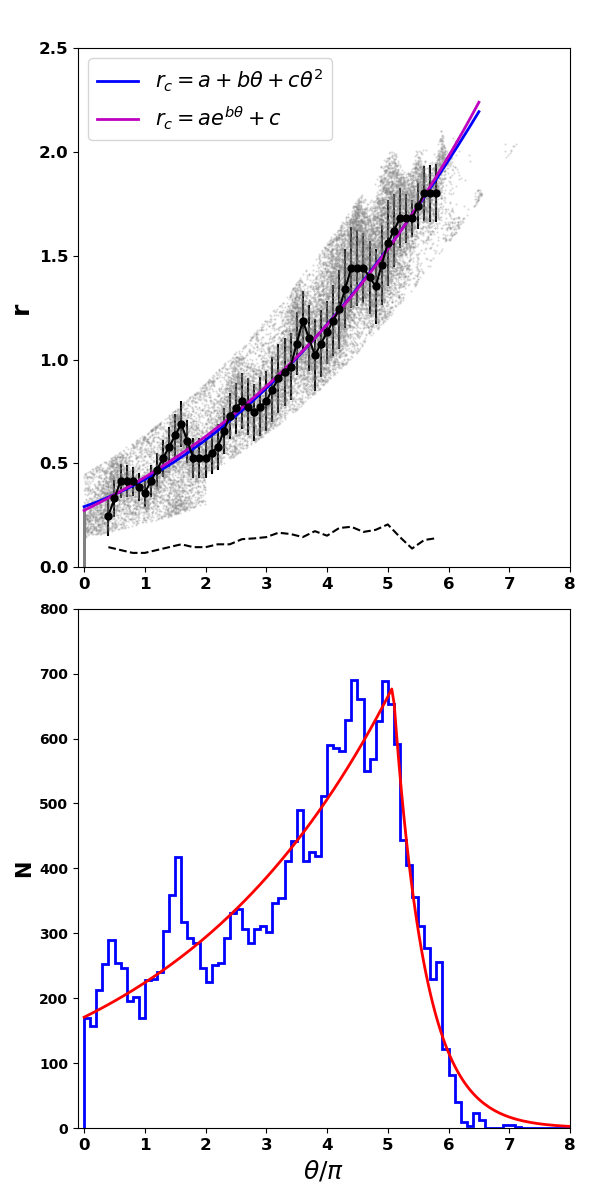}
\caption{
\label{fig:ms_model}
Model fitting to the simulated snail shape (upper panel) and its angular distribution (lower panel). In the upper panel, the grey points show the phase space angles and radii of all the test particles. The phase space radius is calculated from $z$ and $V_{z}$ (i.e. $r=\sqrt{(\frac{z}{1\, {\rm kpc}})^2 + (\frac{V_{z}}{45\, {\rm km/s}})^2}$ ), while the phase space angle $\theta$ is assigned an additional $2k\pi$ according to the nearest wrap of the quadratic spiral (blue line) in the same phase space direction. The black points and colored lines are the same as those in Fig. \ref{fig:ms_2d}. The black dashed line is equal to the error bars of the black points, also indicating the radial width of the snail. In the lower panel, the blue histogram of the phase space angle distribution is fitted by a double exponential function, shown as the red line.}
\end{figure}

The $z-V_{z}$ phase space angles $\theta$ and the scaled radii $r$ of the selected density peaks are plotted in the upper panel of Fig. \ref{fig:ms_model} as the black points. This profile is obviously not a straight line (an Archimedean spiral). Therefore, we fit them with a quadratic and a logarithmic spiral (i.e. equiangular spiral) function, shown as the blue and magenta lines in the upper panel, respectively. They can be well fitted with both functions. The corresponding phase snails of the two functions are also shown in Fig. \ref{fig:ms_2d} on top of the test particles distribution. 

Actually, when we choose the intersections between the snail shape with the $z$ and $V_{z}$ axes from Fig. \ref{fig:ms_2d}, the 6 intersections between the snail and the $z$ axis can be better fitted by these two functions, without any wiggles, than the 7 intersections between the snail and the $V_{z}$ axis. In addition, the ratio of fits with intersections on $z$ and $V_{z}$ axes, i.e. the scale parameter for the velocity, is not the constant 45 $\rm km\,s^{-1}$ as we choose. It changes from 60 $\rm km\, s^{-1}$ in the central part to 40 $\rm km\, s^{-1}$ in the outer region. This variation is related to the shape of the vertical potential, and linked to the fact that the vertical oscillation is not harmonic. The effect of such scale velocity variation could be ignored because the simpler function applied can well describe the shape curve of the phase snail under the current scaling parameters.

With a certain function for the shape curve of the snail, we can assign a phase space angle for each test particle. Each test particle has an initial phase space angle $\theta_{0}$, determined by its $z$ and $V_{z}$ values, in a range of $[0, 2\pi)$. The phase angle increases monotonically along the phase snail from the center to outer regions. In order to determine the phase angle of each particle (i.e., the wrap that it belongs to), we can calculate the relative distances of each particle to the snail shape curve at different wraps, i.e. the points with phase angles of $\theta_{0} + 2k\pi$, where $k$ is an integer. The angle with the smallest distance is chosen as the final phase angle of this particle. With the phase radii and corresponding angles of all the test particles, we can study the radial and angular distributions of all the particles in the snail feature.

The phase space angles and radii of all test particles are shown as grey points in the upper panel of Fig. \ref{fig:ms_model}. The error bars of the measured density peaks, which can roughly indicate the radial width of the snail, are also consistent with the 16\% and 84\% percentiles of phase radii in each phase angle bin. It shows a slightly increasing radial width. The angular distribution of all the test particles is shown in the lower panel of Fig. \ref{fig:ms_model}. Note that the fitting of the snail ignores the central region of the phase space (where the snail is not clear). Thus, the phase radii and angles of particles ($\sim$2 900) that are closest to the origin are all assigned as 0. They are not shown in the angular histogram. 

An interesting finding of this test particle simulation here is the peak at $\theta \sim 5\pi$. An intuitive guess for the angular density distribution of the snail is an exponentially decreasing function. However, considering the peak at $\theta \sim 5\pi$, we also try to fit the angular density distribution with a double exponential function, which decreases from the peak toward both directions. The fit is shown as the red line in the lower panel of Fig. \ref{fig:ms_model}, which roughly fits the angular density distribution. The peak could be due to the initial setup of the vertical perturbation, that is also the perturbed vertical energy distribution. It may also explain the large difference ($\rm \sim 2.6\ km\,s^{-1}$) between the N/S velocity dispersion profiles at $|z| \sim 0.45$ kpc (see the middle panel in a later Fig. \ref{fig:obs_exp}). Therefore, this double exponential distribution profile is adopted in Section \ref{sssec:dexp} to better fit the asymmetric velocity dispersion profiles.

\section{Phase Space Decomposition}
\label{sec:models}
In this work, we attempt to connect the N/S asymmetries with the phase space snail, in order to recover the asymmetric N/S velocity dispersion profiles with an analytical snail model. The basic idea is to decompose the tracer population into an unperturbed (or already phase-mixed), smooth N/S-symmetric background and a perturbed (still in phase-mixing) population resembling a phase space snail. The smooth background is assumed to be in equilibrium, and thus its velocity dispersion profile can be utilized to constrain the local density profile more cleanly. The snail is the result of incomplete phase mixing of the perturbed stars oscillating in the vertical direction. We build analytical models for the shape (i.e. the snail central line) and (radial and angular) density distributions of the snail shell. In this way, we could take the disequilibrium features into consideration and quantitatively model the N/S asymmetries, through fitting the asymmetric N/S velocity dispersion profiles.

\subsection{Model of the background}
\label{ssec:mod_back}
The phase space distribution of the background $f_{\rm bg}(z,V_{z})$ consists of an exponential distribution along $z$ direction and a Gaussian distribution along $V_{z}$ direction, which can be written as:
\begin{equation}
\label{eq:fback}
    f_{\rm bg}(z,V_{z})= \frac{1}{2h_{1}} \exp\left( -\frac{|z|}{h_{1}} \right) \frac{1}{\sqrt{2\pi}\sigma_{z}(|z|)} \exp\left( -\frac{V^{2}_{z}}{2\sigma^{2}_{z}(|z|)} \right)\ ,
\end{equation}
where $h_{1}$ is the scale height of the tracer background, and $\sigma_{z}(|z|)$ is its vertical velocity dispersion profile.

As the smooth background of the tracer is assumed to be in equilibrium, its $\sigma_{z}(|z|)$ can be derived from the Galactic potential $\Phi$ through the vertical Jeans equation. Here, we consider the simplest case, in which the tilt term is ignored. The rotation curve term from the Poisson equation is also neglected, which can be quantified via the Oort constants \citep{Binney2008}. This term has a contribution of about $0 - 0.003$ ${\rm M}_{\odot}\,{\rm pc}^{-3}$ to the local dark matter density \citep{Guo2020}, and can be easily corrected if it is independent of the vertical height.

The total mass density $\rho_{tot}(z)$ consists of an exponential stellar disc, a razor thin gaseous disc and a constant dark matter density $\rho_{\rm dm}$. Thus, integrating the vertical Jeans equation \cite[see the detail of the derivation in][]{Guo2020}, we obtain
\begin{equation}
\sigma_{z}^{2}(|z|)\ =\ g(|z|)\ +\ \frac{\nu(z_{0}) \sigma_{z}^{2}(z_{0}) - \nu(z_{0}) g(z_{0})}{\nu(|z|)}\ ,
\label{eq:sigz}
\end{equation}
where
\begin{equation}
\nu(z) = \int_{-\infty}^{\infty} f_{\rm bg}(z,V_{z}) {\rm d}V_{z} = \frac{1}{2h_{1}} \exp\left( -\frac{|z|}{h_{1}} \right) \ ,
\label{eq:nuz}
\end{equation}
\begin{equation}
  \begin{split}
g(|z|) =\ & 2\pi G h_{1} \left\{\ \Sigma_{\star} \left[\ 1 - \frac{z_{\rm h}}{h_{1} + z_{\rm h}} \exp\left( -\frac{|z|}{z_{\rm h}} \right)\ \right] \right. \\
      &\left. +\ \Sigma_{\rm gas}\ +\ 2 \rho_{\rm dm}(|z| + h_{1})\ \right. \bigg\} \ ,
  \end{split}
\label{eq:fz}
\end{equation}

In Eq. \ref{eq:sigz}, $z_{0}$ and $\sigma_{z}(z_{0})$ are the integration boundary. The former can be arbitrary and is set as 50 pc, and the latter is a free parameter. $\Sigma_{\star}$ and $z_{\rm h}$ in Eq. \ref{eq:fz} are the total surface density and the scale height of the exponential stellar disc. The total gas surface density $\Sigma_{\rm gas}$ is assumed as a constant value of 13.2 ${\rm M}_{\odot}\,{\rm pc}^{-2}$ \citep{Flynn2006}. The scale height of the background in the tracer population is set as $h_{1}= 278.6$ pc \citep{Guo2020} to reduce the degeneracy. The free parameters of the background model are $\mathbf{p_{b}} = (\Sigma_{\star},\ z_{\rm h},\ \rho_{\rm dm},\ \sigma_{z}(z_{0}))$.

\subsection{Model of the snail}
\label{ssec:mod_snail}
Our ctric snail model includes functions for its snail shape curve, density distributions along the snail curve (angular distribution in $\theta$ direction) and perpendicular to the snail (radial distribution in $r$ direction). The phase space distribution of the snail is assumed to be separable in polar coordinates ($r$, $\theta$), i.e. $f_{\rm snail}= f_{s,\theta}(\theta) f_{s,r}(r)$, where $r$ and $\theta$ are the phase space radius and angle.

\subsubsection{Snail shape curve}
\label{sssec:exp}
For the snail shape curve, it is well fitted by an Archimedean spiral in observations \citep[e.g.][]{Bland2019, LiZY2021}. However, in the test particle simulation shown in Section \ref{ssec:tps}, a quadratic or a logarithmic function is required to fit the shape curve of its snail feature. The quadratic spiral has a function of 
\begin{equation}
\label{eq:s_squ}
    r_{c}= a + b\theta\ + c\theta^{2}\ ,
\end{equation}
where $r_{c}$ is the radius of a point on the snail shape curve, $a$, $b$ and $c$ are shape parameters. A snail that starts from the origin requires $a=0$. It will reduce to an Archimedean spiral when $a=0$ and $c=0$ as below: 
\begin{equation}
\label{eq:s_archi}
    r_{c}=  b\theta\ .
\end{equation}

We have applied both the Archimedean and quadratic spiral functions to the observational data. They produce similar results and parameter estimations. In addition, the model predicted parameter $c$ in the quadratic spiral function is about two orders of magnitude smaller than the parameter $b$, which means an Archimedean spiral is good enough to describe the observational snail feature. Therefore, for simplicity, we adopt the Archimedean spiral for the observed snail shell shape in the following analysis.

\subsubsection{Snail radial distribution}
\label{sssec:mod_srd}
Since the snail shell has limited width, at each point on the snail shape curve, we adopt a uniform density distribution perpendicular to the phase snail itself, i.e. along the phase space $r$ direction, covering the width of the phase snail. It has a function of 
\begin{equation}
\label{eq:s_fsr}
    f_{s,r}(r)= \frac{1}{W}\ , \quad |r-r_{c}|< \frac{W}{2}  \ ,
\end{equation}
where $W$ is the width of the snail, and the snail shape curve $r_{c}$ is an Archimedean spiral function of $\theta$ as Eq. \ref{eq:s_archi}.

To avoid negative values of $r$ due to the uniform distribution with a constant width in the most central region, we adjust the width W in the central region as follows:
\begin{equation}
\label{eq:s_dr}
W=
\begin{cases}
2r_{c} & \text{$r_{c} < \frac{\Delta r}{2}$} \\
\Delta r & \text{$r_{c} \geq \frac{\Delta r}{2}$}
\end{cases}
\end{equation}
where $\Delta r$ is the constant width of the snail in outer parts of the $z-V_{z}$ plane.

\subsubsection{Snail angular distribution}
\label{sssec:dexp}
For the angular distribution in $\theta$ direction, an intuitive guess is an exponential function showing as
\begin{equation}
\label{eq:s_fst}
    f_{s,\theta}(\theta)= \frac{1}{\theta_{h}} \exp\left( -\frac{\theta}{\theta_{h}} \right) \ ,
\end{equation}
where $\theta_{h}$ is a scale parameter in $\theta$ direction. 

However, in \cite{Guo2020}, we found that the asymmetric N/S $\sigma_{z}$ profiles show a quite large difference, $\rm \sim 2.6\, km\,s^{-1}$, between N/S at $|z| \sim 0.45$ kpc (also shown in the middle panel of Fig. \ref{fig:obs_dexp}). This difference is much larger than those at smaller and larger vertical heights. In order to explain such differences, also considering the density peak in the angular distribution of the test particles simulation, we build a double exponential angular distribution model. This double exponential angular distribution shows a density peak at a certain angle $\theta_{0}$, and exponentially decreases at smaller or larger $\theta$ values with scale lengths of $\theta_{h1}$ and $\theta_{h2}$, respectively. The angular distribution function of the snail $f_{s,\theta}(\theta)$ is then
\begin{equation}
\label{eq:s_dexp}
f_{s,\theta}(\theta)=
\begin{cases}
N_{0} \exp\left( \frac{\theta - \theta_{0}}{\theta_{h1}} \right) & \text{$0 \leq \theta < \theta_{0}$}\\
N_{0} \exp\left( -\frac{\theta - \theta_{0}}{\theta_{h2}} \right) & \text{$\theta \geq \theta_{0}$}
\end{cases}
\end{equation}
where $N_{0}$ is a normalization parameter,
\begin{equation}
\label{eq:s_dn0}
N_{0}= 1 \bigg/ \left[ \theta_{h1} \left(1- e^{-\frac{\theta_{0}}{\theta_{h1}}} \right) + \theta_{h2} \right] \ .
\end{equation}

With functions of snail shape curve, radial distribution and angular distribution, we can sample stars in polar coordinates in the phase space. Their vertical heights $z$ and velocities $V_{z}$ can be connected to the polar coordinates ($r$, $\theta$) by:
\begin{align}
\label{eq:s_zvz}
    z & = 1\,{\rm kpc} \times r \cos(\theta - \Psi_{0})\ \nonumber\\
    V_{z} & = \upsilon_{z,s} \times r \sin(\theta - \Psi_{0})\ ,
\end{align}
where $\Psi_{0}$ is the current phase angle of the snail in the $z-V_{z}$ plane, and $\upsilon_{z,s}$ is the vertical velocity scale. For this double exponential angular distribution snail model, the free parameters are denoted as $\mathbf{p_{s}} = (f_{s},\ \theta_{h1},\ \theta_{0},\ \theta_{h2},\ b,\ \Psi_{0},\ \Delta r,\ \upsilon_{z,s})$, where $f_{s}$ is the ratio of the stars in the snail feature relative to the background, standing for the fraction of stars which are still phase-mixing.

\subsection{Likelihood and MCMC}
\label{ssec:lnp}
With the analytical models built for the smooth background and the phase space snail, we can then fit the observation with models. As the distribution function for the phase space snail is in polar coordinates, it is complicated to combine the differential probability of a star in $z-V_{z}$ (for the background) and $r-\theta$ (for the snail feature) spaces. Therefore, we take a model sample according to the distribution functions of the background and phase space snail. The total stellar number of the sample is set as $N_{t} = 10^{5}$, of which stars in the background and the snail are $N_{t}/(1+f_{s})$ and $N_{t}*f_{s}/(1+f_{s})$, respectively. For a tracer scale height of 278.6 pc, the number of stars within $0.9< |z| <1.0$ kpc is about 1200 if all the stars are assigned to the background. In addition, $\rm N_{t}$ is enlarged by 5 times in case we have fewer stars in that vertical range.

With the model sampling, we can in principle compare the 2D phase space distributions of the observation and model. However, there are two reasons preventing this. The first one is the complicated selection function due to the LAMOST spectroscopic survey. The selection function is a function of the line-of-sight, distance, magnitude and color. Though we have done the selection effect correction for each star, it actually acts on the integration of the phase space distribution function along $V_{z}$. Thus, the consideration of the selection function will be quite complicated. The second reason is the 4.5 $\rm km\,s^{-1}$ measurement error of velocity. It is convolved into the vertical velocity distribution at a certain vertical height. This measurement error can be considered through randomly drawing realizations for the model velocities, which could be quite time-consuming.

Therefore, we compare only the vertical velocity dispersion ($\sigma_{z}$) profiles with a bin width of 100 pc. This comparison is based on two assumptions: (1) the selection function is independent of the vertical velocity; (2) the selection function in each vertical bin is roughly similar. Though the N/S number density profiles also show the signal of asymmetries, the bootstrapping error of the median value in each vertical bin is small ($\sim 0.01$ dex), while the 16\% and 84\% percentiles errors in each bin is large, $\sim 0.5$ dex. They are not suitable for model comparison. Nevertheless, the velocity measurement error can be easily subtracted from the observed $\sigma_{z}$ profiles.

The log-likelihood is thus given by
\begin{equation}
{\rm ln}\ L = -\, \frac{1}{2}\sum_{i} \left[ \frac{\sigma_{z,i}^{mod} - \sigma_{z,i}^{obs}}{\sigma_{z,i}^{err}} \right]^{2}\ ,
\label{eq:lnp}
\end{equation}
where $i$ indicates the vertical bin and $\sigma_{z,i}^{mod}$ is the velocity dispersion calculated from the model sample. $\sigma_{z,i}^{obs}$ and $\sigma_{z,i}^{err}$ are the observed velocity dispersion and its error, the former of which has been subtracted with the 4.5 $\rm km\,s^{-1}$ measurement error. With this likelihood function, we can utilize the Markov Chain Monte Carlo (MCMC) technique to compare the observed and the model $\sigma_{z}$ profiles and obtain parameters' posterior distributions. The validation of the MCMC being able to properly detect and quantify the numerous parameters of the models is shown in Appendix \ref{ssec:mock}. The MCMC package we use is {\tt EMCEE} \citep{Foreman2013}. We apply flat priors, i.e. wide parameter ranges, for all the parameters, except for a Gaussian prior on $\Sigma_{\star}$ from \cite{Guo2020}.

\section{Results} 
\label{sec:result}

\subsection{Asymmetric velocity dispersion}
\label{ssec:comp_sigz}

\begin{figure*}[bth]
\centering
\includegraphics[width=0.9\textwidth]{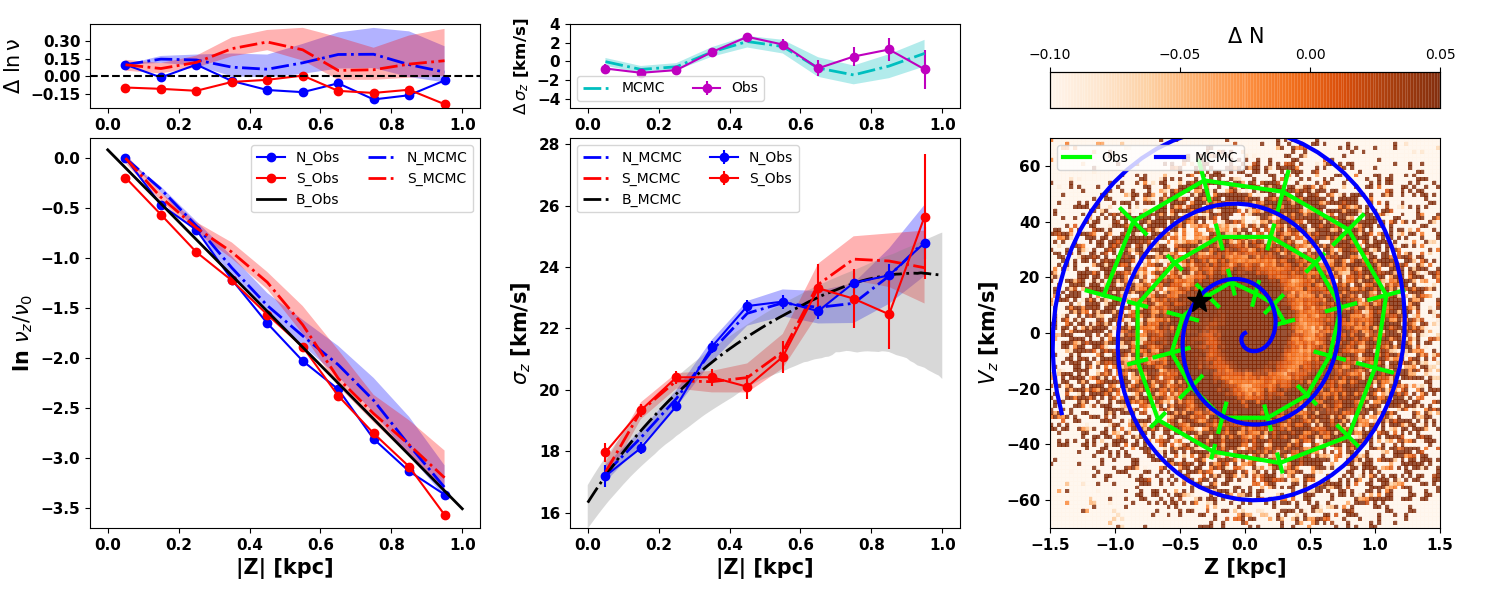}
\caption{
\label{fig:obs_dexp}
Comparisons of model predicted profiles and observations. The left panel and middle panel show the N/S number density and velocity dispersion profiles, respectively. The blue and red solid lines with error bars display the observed N/S profiles, while the blue and red dash-dotted lines with shaded regions show the model predicted asymmetric N/S profiles. The black lines show the profiles of the smooth background. In the left panel, the scale height of the background is set as 278.6 pc. The N/S number density profiles are normalized to the first bin in the North, while the background profile is located between first bins of N/S. The number density error from 16\% and 84\% percentiles in each vertical bin is about 0.5 dex, while the bootstrapping error of the median value is unreasonably small, about 0.01 dex. Thus both are not shown here. The black dash-dotted line with shaded region in the middle panel is the model predicted symmetric $\sigma_{z}$ profiles for the background. Above the number density profiles, we show residuals between the N/S profiles and the smooth background. The small panel above the middel panel shows the differences between N/S velocity dispersion profiles, in which the magenta and cyan lines show the observational and model predicted profiles, respectively. 
The right panel compares the phase snail shapes between the model and observation. The orange map is the density contrast map from \protect \cite{LiZY2021} for stars with the guiding center near the Sun, i.e. $|R_{g} - R_{\odot}| < 0.2$ kpc. The green line with error bars is the phase snail shape curve measured from the number density contrast map, while the blue line is our best-fit model. The black star in the right panel indicates the peak location of the snail angular distribution, derived from the median values of parameters' posterior distributions.}
\end{figure*}

The model fitting to the N/S asymmetric $\sigma_{z}$ profiles and the comparison of the snail shape for our double exponential angular distribution snail model (named as `dexp', shown as Eq. \ref{eq:s_dexp}) are shown in Fig. \ref{fig:obs_dexp}. The model predicted parameters are listed in Tables \ref{tab:bparas} and \ref{tab:sparas}, with the corresponding corner plot shown in Fig. \ref{fig:paras_dexp}. As shown in the middle panels of Fig. \ref{fig:obs_dexp}, our `dexp' model can well recover the asymmetric N/S $\sigma_{z}$ profiles, especially for the large N/S difference at $|z|= 0.45$ kpc. While, for the first bins of N/S profiles (i.e. $|z|< 0.1$ kpc), $\sigma_{z}$ from the model is slightly different from the observation. The slight discrepancy is mainly due to the complicated phase distribution in both the observation and model. The most inner region in $z-V_{z}$ space in observations seems to show a blob without clear snail features, possibly due to the higher density and the more tightly wound snail \citep{Antoja2018, Bland2019, LiZY2021}. The snail width in the inner region of our models has also been modified to avoid negative phase radii. For these two reasons, this discrepancy is expected. The complexity in the inner region can be mitigated in a 2D comparison by a mask function, or by a more physically driven snail model. Nevertheless, this discrepancy will not influence the overall fitting, especially in the region of $0.1< |z| < 0.7$ kpc. Moreover, for the Southern $\sigma_{z}$ profile, the clear discrepancy between the model and observation in $z< -0.7$ kpc results from few stars and the resultant large uncertainties.

The density peak of the snail angular distribution of our `dexp' model can be derived from parameters $(\theta_{0},\ b,\ \Psi_{0},\ \upsilon_{z,s})$. From posterior distributions of these parameters, we obtain the density peak located at $(z,\ V_{z})= (-0.36_{-0.09}^{+0.12}\,{\rm kpc},\ 11.6_{-13.7}^{+5.2}\,{\rm km\,s^{-1}})$. Thus, we note that the observed large difference between N/S in $\sigma_{z}$ at $|z|= 0.45$ kpc may indicate a density peak at this vertical height. We will show that the N/S number density asymmetry is also an evidence of the existence of this density peak in Section \ref{ssec:comp_nuz}.

\begin{table}
\renewcommand\arraystretch{1.3}
\centering
\caption{Posterior distributions of parameters for the background. The bottom two rows are for the `dexp' and `exp' snail models, respectively.}
\label{tab:bparas}
\begin{tabular}{ c | c c c c }
\hline
\hline 
Model & $\Sigma_{\star}$  & $z_{\rm h}$ & $\rho_{\rm dm}$ & $\sigma_{z}(z_{0})$  \\
  & [${\rm M}_{\odot}\,{\rm pc}^{-2}$] & [pc] &  [${\rm M}_{\odot}\,{\rm pc}^{-3}$]  & [$\rm km\,s^{-1}$] \\
  & (1) & (2) & (3) & (4) \\
 \hline
    `dexp'  &  $35.3^{+5.1}_{-5.5}$ & $427.5^{+186.6}_{-145.2}$ & $0.0151^{+0.0050}_{-0.0051}$ & $17.2^{+0.6}_{-0.9}$  \\
   `exp'  &  $37.0^{+5.2}_{-5.5}$ & $247.2^{+58.9}_{-63.4}$ & $0.0096^{+0.0028}_{-0.0040}$ & $17.8^{+0.3}_{-0.3}$ \\
 \hline
\end{tabular}
\begin{tablenotes}
\small
\item Note: Columns are: (1) total surface density of the stellar disc; (2) scale height of the stellar disc; (3) local dark matter density; (4) velocity dispersion at $z_{0}= 50$ pc for the background $\sigma_{z}$ profile.
\end{tablenotes}
\end{table}

\begin{table*}
\renewcommand\arraystretch{1.3}
\centering
\caption{Posterior distributions of parameters for the phase space snail. The rows are the same as Table \ref{tab:bparas}.}
\label{tab:sparas}
\begin{tabular}{ c | c c c c c c c c c}
\hline
\hline
Model & $f_{s}$   & $\theta_{h}$   & $\theta_{h1}$   & $\theta_{0}$ & $\theta_{h2}$ &    b   &  $\Psi_{0}$   &  $\Delta r$    & $\upsilon_{z,s}$ \\
  & [rad] & [rad] & [rad] & [rad] & & [rad] & & [$\rm km\,s^{-1}$] \\
  & (5) & (6) & (7) & (8) & (9) & (10) & (11) & (12) & (13) \\
\hline
  `dexp'  & $0.161^{+0.125}_{-0.047}$ & -- & $4.98^{+3.45}_{-2.22}$ & $5.39^{+1.02}_{-0.61}$ & $2.79^{+1.87}_{-0.92}$ & $0.0799^{+0.0094}_{-0.0085}$ & $2.78^{+0.93}_{-0.65}$ & $0.27^{+0.09}_{-0.06}$ & $54.4^{+4.2}_{-8.2}$  \\
 `exp'  & $0.133^{+0.038}_{-0.031}$ & $3.94^{+1.09}_{-0.65}$ & -- & -- & -- & $0.0961^{+0.0102}_{-0.0059}$ & $1.84^{+0.25}_{-0.32}$ & $0.26^{+0.06}_{-0.05}$ & $48.0^{+5.6}_{-4.8}$ \\
 \hline
\end{tabular}
\begin{tablenotes}
\small
\item Note: Columns are: (5) ratio of the snail relative to the background; (6) scale height of the single exponential angular distribution; (7) inner scale height of the double exponential angular distribution; (8) phase angle of the density peak; (9) outer scale height of the double exponential angular distribution; (10) coefficient of the Archimedean spiral; (11) phase angle of the snail; (12) radial width of the snail; (13) scale velocity.
\end{tablenotes}
\end{table*}

\subsection{Comparisons of the snail shapes}
\label{ssec:comp_shape}
The comparison of the model predicted snail shape curve with the observation is shown in the right panel of Fig. \ref{fig:obs_dexp}. The background density contrast map is taken from \cite{LiZY2021}, for the observed snail shape at $R_{g}= 8.34\pm0.2$ kpc in the Solar Neighborhood. The number density contrast is derived as $\Delta N= N/\widetilde{N} -1$, where $N$ and $\widetilde{N}$ are the observed and the Gaussian kernel convolved number densities in each phase space bin, respectively \citep{Laporte2019, LiZY2020}. The $R$- and $R_{g}$- based phase space snails look similar, while the snail for the latter is clearer and shows one more wrap \citep[see Figs. 10, 11 in][]{LiZY2021}. The snail shape can be extracted from the density contrast map by identifying the peaks along the snail, which is displayed as the green curve in the right panel. The Archimedean spiral function fitting of this observational snail shape curve can be presented by snail parameters as $(b,\ \Psi_{0},\ \upsilon_{z,s}) = (\frac{1}{6\pi},\ 0.,\ 46.7\ {\rm km\,s}^{-1})$.

The snail shape curves from the observation and our model are consistent within a phase radius range of $0.2< r <0.7$ (roughly the range of $0.2< z/kpc <0.7$). In the inner region of $r <0.2$, the difference shows in the obtained values of $\Psi_{0}$, which are 0 and 2.78 for the observation and the `dexp' model, respectively. In the outer region of $r >0.7$, the different snail shape curves become divergent. It is also indicated by the different values of the Archimedean spiral coefficient $b$. The model predicted value ($b= 0.0799^{+0.0094}_{-0.0085}$) is larger than that from the observation ($b= 0.0531$). There are two reasons for the poor constraint on the parameter $b$ and the snail shape in the outer region. The first reason is the one dimensional comparison between the model and observation, which is applied to the $\sigma_{z}$ profiles. For both the model and observation, there are more stars of the snail feature in the inner region than the outer region. Therefore, the one dimensional comparison is dominated by the inner region, where the snail shape is roughly well constrained. This situation will be alleviated for a 2D comparison, especially when the inner region is masked. The second reason could be the noise in the outer part. In the observed density contrast map, the snail shape in the outer region ($r \gtrsim 0.7$) is not very clear and could be more noisy when the stars are radially binned by Galactocentric radius $R$, than the snail shape from stars binned by guiding radius $R_{g}$. In addition, $\sigma_{z}$ profiles of our sample show larger uncertainties in the outer region, especially for the Southern sky.

\begin{figure}[bth]
\centering
\includegraphics[width=0.9\columnwidth]{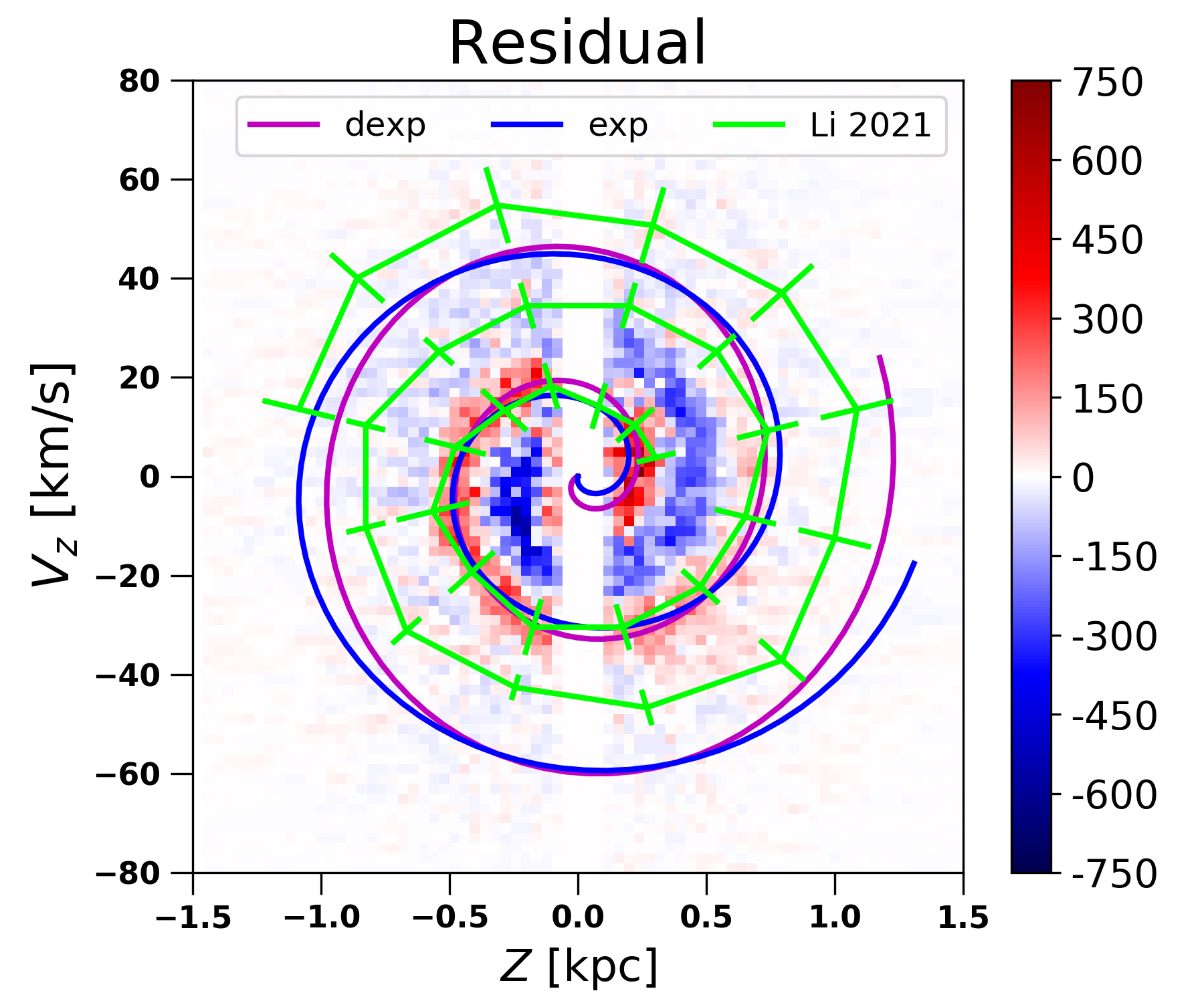}
\caption{
\label{fig:obs_comp}
Comparison of the snail shapes with the residual map from \protect \cite{LiH2021}. The green curve with error bars is the snail shape from the density contrast map \protect \citep{LiZY2021}, while the magenta and blue lines are snail shape curves from our `dexp' and `exp' snail models, respectively.}
\end{figure}

In Fig. \ref{fig:obs_comp}, we show the comparison of the snail shapes from our model with that shown in the residual map from \cite{LiH2021}, and the observation result in \cite{LiZY2021}. In \cite{LiH2021}, the authors found a clear snail feature in the residual map between the phase space distributions from the observation and their model. In their model, the tracer is assumed to have a Rational Linear Distribution Function, which can be represented as a continuous superposition of isothermal distribution functions. As shown in Fig. \ref{fig:obs_comp}, the snail shape curve from our `dexp' model fits best to the residual map in the region of $|z|< 0.7$ kpc, even showing a slightly better consistency than that between the residual map and the density contrast map from \cite{LiZY2021}.

\subsection{N/S number density asymmetry}
\label{ssec:comp_nuz}
The model predicted N/S number density profiles for our `dexp' snail model are shown in the left panels of Fig. \ref{fig:obs_dexp}. The N/S difference pattern, i.e. rising in turns, is similar between the observation and model. However, it shows some differences between the profiles from the observation and model. The intersections of N/S number density profiles are also slightly shifted. These differences could be due to the large dispersion ($\sim$0.5 dex) in number densities derived from the selection effect correction for each star in each vertical bin.

\begin{figure}[bth]
\centering
\includegraphics[width=0.9\columnwidth]{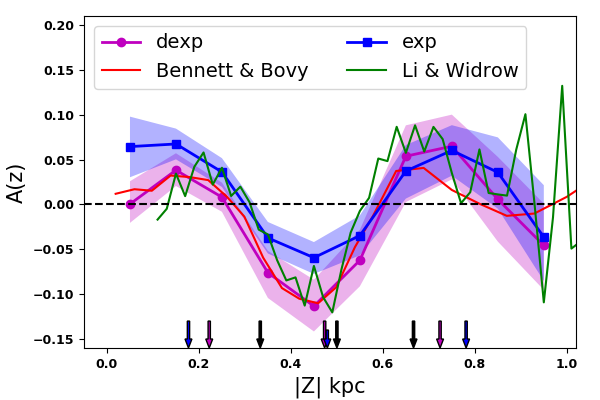}
\caption{
\label{fig:nu_NS}
The N/S vertical number density difference profile. The magenta and blue lines are for our `dexp' and `exp' models, respectively. The shaded regions are the $1\sigma$ error bars from MCMC. The red and green lines are the median profile of samples with different stellar colors, from \protect \cite{Bennett2019}, and the profile of a giant star sample from \cite{LiH2021}. The colored arrows show the intersections between the different snail models with the $z$ axis. The black arrows are derived from the observed snail shape \protect \citep{LiZY2021}, while others are from our snail models.}
\end{figure}

The N/S number density asymmetries have been observed in previous works \cite[e.g.][]{Widrow2012, Bennett2019}. In these works, the N/S number density difference is quantified as: 
\begin{equation}
\label{eq:Anuz}
    A(z)= (\nu(z)-\nu(-z))/(\nu(z)+\nu(-z))\ ,
\end{equation}
where $\nu(z)$ is the number density at vertical height z. This N/S difference is found to be independent of the stellar color. We can also derive the $A(z)$ profile for our snail model. The comparison is shown in Fig. \ref{fig:nu_NS}. Both the median profile from \cite{Bennett2019} and the $A(z)$ profile of a giant sample from \cite{LiH2021} show clear peaks at $|z| \approx 0.2,\ 0.7$ kpc, and a clear deficit at $|z| \approx 0.45$ kpc. 

The $A(z)$ profile from our `dexp' snail model, shown as the magenta line in Fig. \ref{fig:nu_NS}, is well consistent with previous works, especially for the locations of deficits and peaks. Our model even shows high consistency at amplitudes of $A(z)$ profile. Note that, our `dexp' model is derived only by utilizing the vertical velocity dispersion profile. The good consistency is a strong indication of a tight connection between the N/S asymmetries in the kinematics, number density and the phase space snail feature.

\subsection{Vertical bulk motion}
\label{ssec:comp_vz}
\begin{figure}[bth]
\centering
\includegraphics[width=0.9\columnwidth]{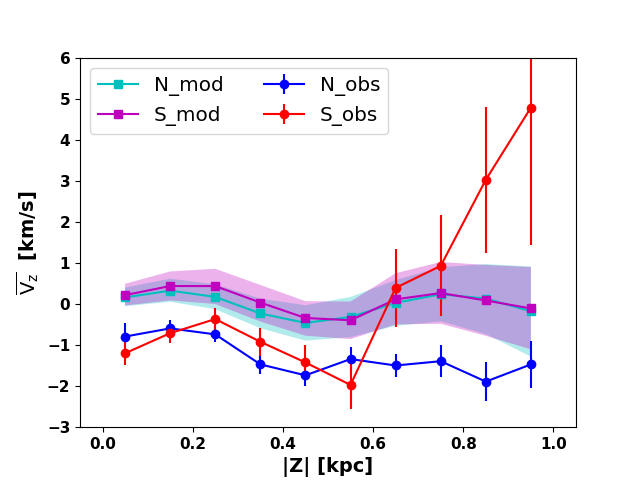}
\caption{
\label{fig:vz_NS}
The vertical bulk motions in N/S. The blue and red lines represent the observed Northern and Southern mean vertical velocities, respectively. The cyan and magenta lines are both derived from our `dexp' model, for the North and South, respectively. The shaded regions are 1$\sigma$ errors.}
\end{figure}

The vertical bulk motion is also an indicator of the disequilibrium in the Milky Way disc. It has been extensively studied and discussed in previous works \citep{Widrow2012, Carlin2013, Williams2013, SunNC2015, Carrillo2018, Gaia2018a, Bennett2019}. The recent works support a combination of breathing and bending modes for the Galactic disc \citep{Carrillo2018, Gaia2018a}, and a downward bending mode is found in the Solar vicinity ($R\approx R_{\odot},\ |z|< 2$ kpc) \citep{Gaia2018a, Bennett2019}. As shown in Fig. \ref{fig:vz_NS}, our dwarf sample also displays a clear downward bending mode at $|z|< 0.5$ kpc, with an amplitude of about $\rm 1.5\ km\,s^{-1}$. The higher region seems to show a breathing mode (compression), which could be less certain due to the large uncertainties in the South.

The phase space snail will result in the non-zero mean vertical velocity profile for the model snail, as the symmetric background has no vertical bulk motion. The vertical bulk motion of our `dexp' model is shown in Fig. \ref{fig:vz_NS}. There are several interesting findings in this model predicted bulk motion. First, the N/S mean vertical velocities are similar (bending mode) but very close to zero, as the snail is roughly symmetric about the $z$ and $V_{z}$ axes (considering the snail having a roughly circular shape). Second, the N/S bulk motions do display a wave-like pattern, though the amplitude is only about 1 $\rm km\,s^{-1}$. Finally, the observed mean vertical velocity profiles seem to be vertically shifted from the model profiles, when ignoring the four outer bins with large uncertainties in the South. The vertical shift is about 1.2 $\rm km\,s^{-1}$. This shift could be partially due to the Solar vertical velocity ($\rm W_{\odot}$) applied. As shown in Fig. 9 of \cite{TianH2015}, the Solar motion shows a correlation with the effective temperature ($T_{\rm eff}$) of the sample. A sample with $T_{\rm eff} \sim 5500$ K, which is the case for our dwarf sample, gives an 1 $\rm km\,s^{-1}$ smaller $\rm W_{\odot}$ than the usual quoted value of $\sim 7\ {\rm km\,s^{-1}}$. We cannot discard other possibilities, such as a warp in the disc. Nevertheless, it will not influence the model comparison of the velocity dispersion. More studies about the connection between the vertical bulk motion and the phase space snail are needed.

\subsection{The smooth background}
\label{ssec:comp_back}
The background in our snail model is assumed to be in equilibrium, and its $\sigma_{z}$ profile is derived from the Galactic mass models. The parameters for the background $\sigma_{z}$ profile are listed in Table \ref{tab:bparas}. The local dark matter density derived from our `dexp' snail model is $0.0151^{+0.0050}_{-0.0051}$ ${\rm M}_{\odot}\,{\rm pc}^{-3}$. This value is consistent with the value of $0.0133^{+0.0024}_{-0.0022}$ ${\rm M}_{\odot}\,{\rm pc}^{-3}$ in our previous work \cite{Guo2020}. Nevertheless, it has larger error bars, though we used the same Gaussian prior on the stellar surface density $\Sigma_{\star}$. Its larger error bars make it consistent with many local measurements utilizing the vertical kinematics, such as \cite{Bienayme2014}, \cite{Xia2016}, \cite{Hagen2018}, \cite{Sivertsson2018}, \cite{Salomon2020}. It is still larger than most global measurements by constraining the global Milky Way potential with the rotation curve, e.g. \cite{Huang2016}, \cite{Salas2019}, \cite{Cautun2020}, and than some other local measurements \citep[e.g.][]{Bovy2013, Zhang2013}. See \cite{Salas2020} for a review of recent measurements.

The large uncertainty of model parameters results from the degeneracy between the snail and the background in the fitting with only $\sigma_{z}$ profiles, as well as the degeneracy between the baryon and dark matter in constructing the background potential. The latter probably contributes more, as shown by the stronger degeneracy between the stellar surface density and the local dark matter density in the corner plots of Fig. \ref{fig:paras_dexp}. As our data covers a small vertical range of $|z| <1.0$ kpc, and the model takes more weights in an even smaller range of $|z| <0.8$ kpc, the degeneracy between the baryon and dark matter would be very large as illustrated in Fig. 14 of our previous work \cite{Guo2020}.

\begin{figure}[bth]
\centering
\includegraphics[width=0.9\columnwidth]{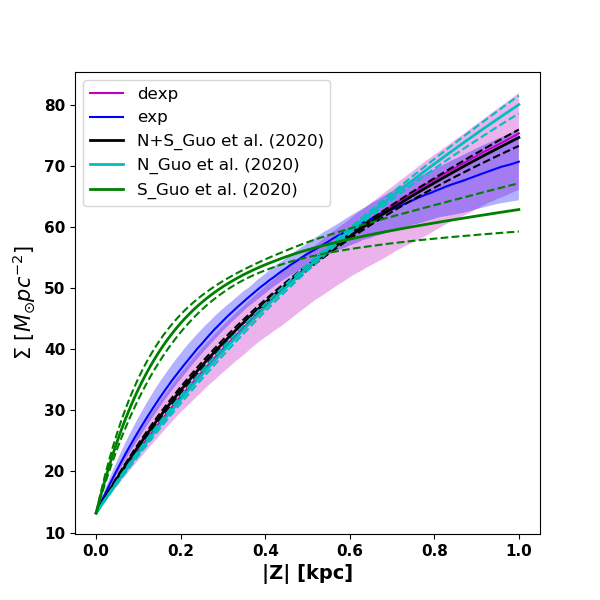}
\caption{
\label{fig:ss_NS}
The total surface density profiles from the background $\sigma_{z}$ profiles for our `dexp' (magenta) and `exp' (blue) snail models. The cyan, green and black lines are taken from \protect \cite{Guo2020}, for samples in the North, South and the combined, respectively. In their work, samples are assumed to be in equilibrium. The shaded regions and dashed lines are their 1$\sigma$ errors.}
\end{figure}

Fig. \ref{fig:ss_NS} compares the total surface mass density between the model and our previous results \citep{Guo2020}. The total surface densities are also roughly consistent at large vertical height. The typical uncertainty of $\sim$5-10 ${\rm M}_{\odot}\,{\rm pc}^{-2}$ is much larger than that from \cite{Guo2020}. The difference at $|z|< 0.6$ kpc results from the slight difference in the model predicted $z_{h}$, which usually has a quite large uncertainty. 

In our previous work of \cite{Guo2020}, which is a typical equilibrium dynamical modelling, we found the N/S subsamples resulted in quite different and inconsistent $\rho_{\rm dm}$. This is mainly due to the large N/S difference at $|z| \sim 0.45$ kpc. Though after considering the N/S difference as an unknown systematic error, we could obtain more consistent $\rho_{\rm dm}$ values. However, we could not fit the N/S $\sigma_{z}$ profiles simultaneously. This is also shown by the cyan and green lines in Fig. \ref{fig:ss_NS}, which are different surface density profiles predicted using samples in the North and South respectively. In this work, we simultaneously fit well the asymmetric N/S $\sigma_{z}$ profiles through the phase space decomposition. The large uncertainty in the background potential can be reduced by 2D phase space ($z-V_{z}$ space) comparison for a sample with simpler selection effects. Besides, in the future, we can apply priors derived from observations for the snail parameters, to reduce the degeneracy between the snail and background.

In the more recent works of \cite{Widmark2021a, Widmark2021b}, the phase space snail is also considered in their dynamical models. They have built models for the `bulk' phase space density and the phase space snail. However, their `bulk' distribution is different from our background. Their `bulk' phase space density is modelled by several bivariate Gaussians, while our background distribution is determined by the Galactic potential. In their model, the snail is considered as a relative number density with respect to the `bulk', the amplitude of which changes with the vertical energy and the perturbation evolution time. Thus the modelling of the snail starts from the perturbation and is not separable from the `bulk' density distribution. Their snail feature shows more clearly in the relative difference between the data and the `bulk', which is similar to the residual map of \cite{LiH2021}. The modelling of the snail by the phase space orbit integration is helpful to build a more physical snail model.

\section{Discussion} 
\label{sec:dis}
\subsection{Density distribution along the snail width}
The distribution of stars perpendicular to the snail shape curve in our snail model is assumed to be uniform with an almost constant phase space radial width as shown in Eq. \ref{eq:s_dr}. We have also tested a Gaussian radial distribution (in which the width $\Delta r$ is the Gaussian dispersion). Most of posterior distributions of the model parameters are consistent with those of the uniform radial distribution, except for a relatively large snail fraction $f_{s} = 0.443 \pm 0.062$ emerging in the `dexp' model.

The snail shape from the observed density contrast map is derived from the difference between the original and the Gaussian kernel convolved number density distributions. The convolution is influenced by the distribution function of the background, the snail fraction and also the radial distribution of the snail. A snail with an increasing distribution perpendicular to the snail shape curve will result in a different density contrast map, compared to a snail having a decreasing distribution. The snail shape from our `dexp' model has a better consistency with the residual map from \cite{LiH2021} than that with the density contrast map from \cite{LiZY2021} in the inner phase space (r< 0.4), as shown in Fig. \ref{fig:obs_comp}. This could be partly due to the radial distribution function of the snail and the convolution used in the density contrast map.

The density distribution along the snail width may be related to the azimuthal velocity. The phase space snail is found to be clearer when color-coded by the azimuthal velocity, which indicates a strong correlation between the in-plane and vertical motions \citep{Antoja2018, Laporte2019}. In the test particle simulations of \cite{LiZY2021}, the snail patterns show a radial dependence on the azimuthal velocity when the particles are binned by the Galactocentric radius, as shown in their Figs. 18 and 20. In contrast, when the particles are binned by the guiding center radius, there is no systematic variation of the snail shape with the azimuthal velocity. Actually, the azimuthal velocities of particles at a given radius are proportional to the angular momentums, which also indicate the particles' guiding center radii. The radial dependence of the snail shape on the azimuthal velocity is due to the mixture of stars with different angular momentum, i.e. different guiding center radii \cite[see][]{LiZY2021}. Therefore, in the future, perhaps we can utilize the azimuthal velocity as an additional parameter to help us construct the radial distribution of the snail model.

\subsection{A different snail model: single exponential angular distribution}
\begin{figure*}[bth]
\centering
\includegraphics[width=0.9\textwidth]{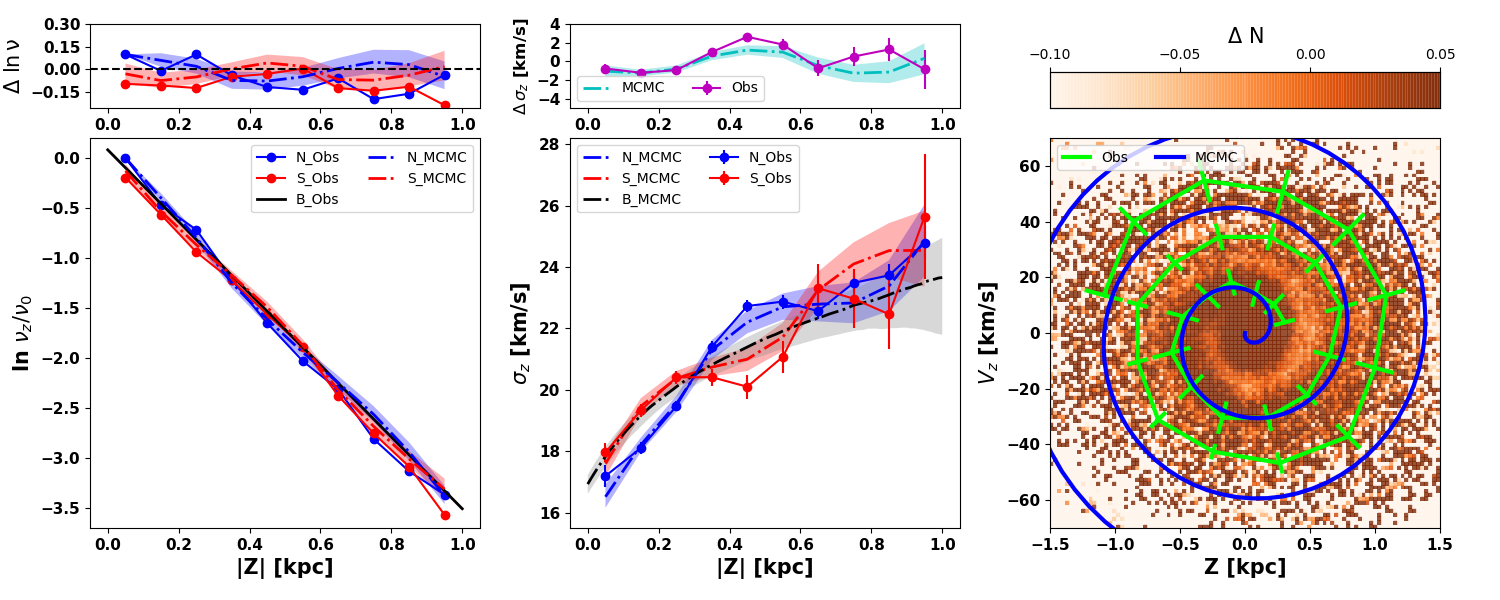}
\caption{
\label{fig:obs_exp}
The same figure as Fig. \ref{fig:obs_dexp} but for the snail model having a single exponential angular distribution function.}
\end{figure*}

We have also tried a snail model with an intuitive single exponential angular distribution (labeled as `exp', shown in Eq. \ref{eq:s_fst}), inspired by the exponential vertical number density. This model has two fewer parameters than the `dexp' model. The results are shown in Fig. \ref{fig:obs_exp}. The `exp' model can also roughly recover the asymmetric N/S $\sigma_{z}$ profiles, as shown in the middle panels of Fig. \ref{fig:obs_exp}. However, the N/S difference in $\sigma_{z}$ from this model is not as large as that in the observation at $|z|= 0.45$ kpc, where the value in the South is obviously larger than that from the observation.

In our snail models, we have four free parameters for the smooth background, and 8 (6) parameters for the `dexp' (`exp') phase space snail. However, we only have 20 data points for the N/S velocity dispersion profiles when the vertical bin size is $\Delta z= 100$ pc. Thus the degeneracy in model parameters is quite large, which results in the large uncertainties in the model predicted posterior distributions, even though a Gaussian prior on $\Sigma_{\star}$ is applied. We have also tried a smaller bin size $\Delta z= 50$ pc, under which we have 40 data points. The results are consistent with those from the bin size of $\Delta z= 100$ pc within error bars.

For all the results using a uniform or a Gaussian phase space radial distribution, or a smaller bin size, $(z_{\rm h},\ \sigma_{z}(z_{0}),\ \theta_{h},\ \theta_{0},\ b,\ \Psi_{0})$ are similar for the `dexp' or `exp' snail model. In contrast, $(z_{\rm h},\ f_{s},\ \Psi_{0})$ from the `dexp' model are larger than those from the `exp' model, while $(b,\ \sigma_{z}(z_{0}))$ are slightly smaller. There are several reasons for these parameter differences.

Firstly, the critical constraint in the `dexp' model is the location of the density peak in the snail, which is resulted from the large $\sigma_{z}$ difference at $|z| \sim 0.45$ kpc. For all the results from the `dexp' snail model, the constraints on the phase angle of the density peak, i.e. $\theta_{0} - \Psi_{0}$, are quite consistent with each other. The locations of the density peak from the Gaussian phase space radial distribution and $\Delta z= 50$ pc are $(z,\ V_{z})= (-0.44_{-0.05}^{+0.12}\,{\rm kpc},\ 3.6_{-12.3}^{+9.3}\,{\rm km\,s^{-1}})$ and $(z,\ V_{z})= (-0.38_{-0.08}^{+0.10}\,{\rm kpc},\ 9.5_{-9.1}^{+5.9}\,{\rm km\,s^{-1}})$, respectively.

Secondly, the intersections between the snail and the $z$ axis, i.e. $z= b(\Psi_{0} + k\pi)$ have strong constraints on $b$ and $\Psi_{0}$. For the `dexp' and `exp' snail models, the phase angle of the snail $\Psi_{0}$ is $\sim 2.8$ and $\sim 2.0$. The snail starts in the third quadrant for the `exp' model, while the `dexp' model prefers a larger $\Psi_{0}$ that is closer to the second quadrant. Therefore, the `exp' snail model will obtain a smaller $b$ than the `dexp' model.

Finally, the angular profile of the `dexp' model from the origin to $\theta_{0}$ is an exponentially increasing function, which differs greatly from the exponentially decreasing profile of the `exp' model. This results in that the `exp' model will have many more stars in the North at the first two bins ($|z|< 0.2$ kpc). Therefore, for the `exp' model, $\sigma_{z}$ in the North is much smaller than that in the South, and $\sigma_{z}(z_{0})$ of the background needs to be closer to that value in the South. It then results in a larger $\sigma_{z}(z_{0})$ for the `exp' model than the `dexp' model. Under the Gaussian prior of $\Sigma_{\star}$, a larger $\sigma_{z}(z_{0})$ will lead to a smaller scale height for the stellar disc. In addition, the larger snail fraction $f_{s}$ for the `dexp' snail model is inherent in the requirement of fitting the larger $\sigma_{z}$ difference at $|z| \sim 0.45$ kpc.

In summary, the large difference of $\sigma_{z}$ at $|z| \sim 0.45$ kpc is a signal of a density peak of the snail feature, and can result in a strong constraint on the location of that density peak. The angular distribution of the observed snail could be a combination of the `dexp' and `exp' models, highly depending on the perturbation model. The coefficient $b$ of the Archimedean spiral is tightly constrained by the intersections of the snail feature with the $z$ axis. The integral boundary in the Jeans equation, i.e. $\sigma_{z}(z_{0})$, is highly influenced by the snail model. It is also influenced by the snail in the very central region, which is not considered in our current models. A snail model built from the perturbation, such as the impulse approximation, will be more realistic than the analytical snail model applied here. In addition, 2D comparison between the model and observation will give better constraint on the snail feature, especially for the outer part when we mask out the inner region. This will be dealt with in a future work.

\subsection{Snail fraction and the influence of self-gravity}
The basic assumption in our models is that the background and the snail feature are separable. The background is either not affected by the perturbation, or has finished the phase mixing. Thus, only the stars in the snail feature are perturbed, such as being imposed a vertical velocity kick. In addition, the interaction between the background and the snail is not considered, which means that the self-gravity is ignored. This assumption is quite similar to \cite{LiH2021} when they generated their partially perturbed mock data set. It is also the reason that the snail shape from our `dexp' model is quite consistent with the residual map from \cite{LiH2021}.

In our snail models, the snail fraction $f_{s}$ relative to the background is also predicted by the model comparison, which means that the fraction of stars still in phase-mixing relative to the total tracer population is $f_{s}/(1+f_{s})$. The values are listed in Table \ref{tab:sparas}. The fraction from the `exp' model is about 0.133, which is slightly smaller than the value ($\sim$ 0.161) from the `dexp' model. It means that about 13.9\% (11.7\%) stars in the sample are still in phase-mixing, for the `dexp' (`exp') snail model. This fraction is even higher for the model with a Gaussian radial distribution. As a result, the self-gravity, i.e. the potential associated with the perturbation acting on the unperturbed disc may be non-negligible.

As discussed in \cite{Darling2019}, when self-gravity is included, the disc develops persistent vertical oscillations. The bending waves persist with roughly constant amplitudes for 1 Gyr, though the snail patterns that arise when including self-gravity are not as well defined as those with test particles. Proper consideration of the self-gravity is very complicated in analytical models and time-consuming in simulations, which is beyond the scope of this work.

\section{Conclusion}
\label{sec:con}
In order to better understand the origin of the asymmetric N/S velocity dispersion profiles, we take the phase space snail into consideration in our model construction. We decompose the tracer population into an unperturbed (or fully phase-mixed) smooth background and a perturbed (still in phase-mixing) phase space snail. The smooth background is assumed to be in equilibrium and N/S symmetric. Thus its vertical velocity dispersion profile is determined by the Milky Way potential through the vertical Jeans equation, and the observed asymmetric N/S velocity dispersion is mainly contributed by the phase snail. The phase space snail is constructed by the snail shape curve (an Archimedean spiral), with additional considerations on the phase space radial (uniform) and angular distributions. The 1D velocity dispersion profiles are utilized to constrain the model parameters with the MCMC technique.

Inspired by the test particle simulation, we build a snail model, i.e. the `dexp' model, with a double exponential angular distribution. It shows a density peak at a certain phase angle. This model can fit well the N/S $\sigma_{z}$ profiles at $0.2< |z| < 0.6$ kpc. The snail shape from our `dexp' model is consistent with that from the density contrast map of \cite{LiZY2021} in the inner region, and is even more consistent with the residual map from \cite{LiH2021}. The outer region of the phase space shows large discrepancies due to the noise and smaller weights in the fitting. In addition, the N/S number density difference derived from our `dexp' model is consistent with that from previous observations. This indicates a strong correlation between the phase space snail and the N/S asymmetries (including both the number density and kinematics).

Our model construction strategy works well in practice, though our geometric snail model is still simple. Actually, the number density enhancement to the background due to the phase space snail also influences the vertical profiles of the radial and azimuthal velocity dispersions. These can help to build more complicated snail models, such as the dependence of the snail shape on the azimuthal velocity. A more physically motivated snail model can be constructed starting from the modelling of the perturbation. In the future, 2D phase space comparison may be helpful to better constrain the snail shape, especially in the outer region. A sample with simpler selection effects will be utilized for these studies.

\acknowledgments
We thank Haochuan Li \& Lawrence M. Widrow for kindly providing the residual map for the comparison of the snail shape. The research presented here is partially supported by the National Key R\&D Program of China under grant No. 2018YFA0404501; by the National Natural Science Foundation of China under grant Nos. 11773052, 11761131016, 11333003, 12025302, 12122301, 12103031; and by the ``111'' Project of the Ministry of Education under grant No. B20019 and by the Natural Science Foundation of Shanghai (No. 21ZR1430600). We acknowledge the science research grants from the China Manned Space Project with No. CMS-CSST-2021-B03. RG is supported by Initiative Postdocs Supporting Program (No. BX2021183), funded by China Postdoctoral Science Foundation. J.S. acknowledges support from a Newton Advanced Fellowship awarded by the Royal Society and the Newton Fund. ZYL acknowledges support from the Cultivation Project for LAMOST Scientific Payoff and Research Achievement of CAMS-CAS. This work made use of the Gravity Supercomputer at the Department of Astronomy, Shanghai Jiao Tong University, and the facilities of the Center for High Performance Computing at Shanghai Astronomical Observatory. This work made use of the facilities of the Center for High Performance Computing at Shanghai Astronomical Observatory.

Guoshoujing Telescope (the Large Sky Area Multi-Object Fiber Spectroscopic Telescope LAMOST) is a National Major Scientific Project built by the Chinese Academy of Sciences. Funding for the project has been provided by the National Development and Reform Commission. LAMOST is operated and managed by the National Astronomical Observatories, Chinese Academy of Sciences. This work has made use of data from the European Space Agency (ESA) mission {\it Gaia} (\url{https://www.cosmos.esa.int/gaia}), processed by the {\it Gaia} Data Processing and Analysis Consortium (DPAC, \url{https://www.cosmos.esa.int/web/gaia/dpac/consortium}). Funding for the DPAC has been provided by national institutions, in particular the institutions participating in the {\it Gaia} Multilateral Agreement.

\appendix
\section{Mock Test}
\label{ssec:mock}
To validate that the procedure of sampling and MCMC fitting can properly detect and quantify the numerous parameters of the models, we make a mock simulation. For simplicity, this mock data set is not built by a test particle simulation, but by a phase space snail superimposed on a smooth background. The models are the same as Section \ref{sec:models}, except for the velocity dispersion profile of the background. The $\sigma_{z}(|z|)$ profile of the background is given by a linear function of vertical height $|z|$:
\begin{equation}
\label{eq:mock}
\sigma_{z}(|z|)= k_{z} |z| + \sigma_{z,0}\ ,\\
\ k_{z}= 10\ {\rm km/s/kpc}\ ,\ \sigma_{z,0}= 16\ {\rm km/s} \ .
\end{equation}

The phase space snail has an exponential decreasing angular profile, which starts from $\theta= \pi$, rather than the origin ($\theta=0$). This setting is applied to obtain more similar $\sigma_{z}$ as the observation in this work. The scale height ($h_{1}$) of the background is left as a free parameter, while the scale velocity $\upsilon_{z,s}$ is set as 55 $\rm km\,s^{-1}$. Therefore, the initial inputs for the background are $h_{1}= 300\ {\rm pc},\ k_{z}= 10\ {\rm km\,s^{-1}\,kpc^{-1}}\ ,\ \sigma_{z,0}= 16\ {\rm km\,s^{-1}}$, and the inputs for the phase space snail are $f_{s}= 0.2,\ \theta_{h}=4.71,\ b= 0.0637,\ \Psi_{0}=0,\ \Delta r=0.2$ (see Section \ref{ssec:mod_snail} for the meanings of these symbols). The numbers of stars in the background and the phase space snail are 80,000 and 16,000, having a total number comparable to our dwarf star sample.

\begin{figure*}[bth]
    \centering
    \includegraphics[width=0.9\textwidth]{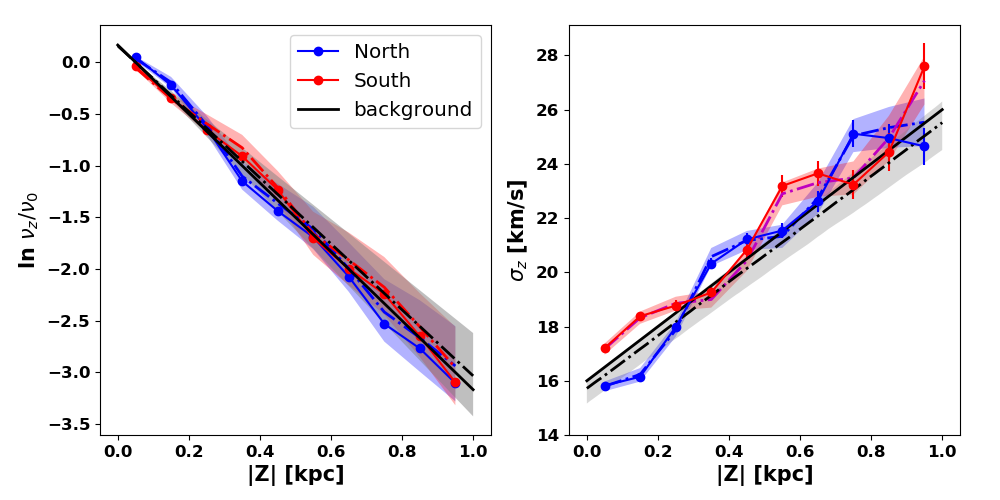}
    \caption{
    \label{fig:mock}
    Number density (left) and velocity dispersion (right) profiles of the mock data (solid lines) and the model predicted ones (dash-dotted lines). The blue and red lines show the Northern and Southern profiles, respectively. The black lines indicate the profiles of the symmetric and smooth background. The shaded regions represent the error bars predicted by MCMC. }
\end{figure*}

The comparison of the number density and velocity dispersion profiles of the mock data and the ones predicted from MCMC are shown in Fig. \ref{fig:mock}. The model obtains N/S profiles which are well consistent with the inputs. The posterior distributions of the parameters from MCMC are $h_{1}= 313\pm40\ {\rm pc},\ k_{z}= 9.8\pm0.7\ {\rm km\,s^{-1}\,kpc^{-1}}\ ,\ \sigma_{z,0}= 15.7\pm0.4\ {\rm km\,s^{-1}}$ and $f_{s}= 0.251\pm0.044,\ \theta_{h}=5.29\pm0.69,\ b= 0.0614\pm0.0025,\ \Psi_{0}=-0.021\pm0.207,\ \Delta r=0.205\pm0.033$. The model predicted values are well consistent with the inputs. This mock test demonstrates that the decomposition, even when only the 1D $\sigma_{z}$ profiles are utilized for the model comparison, is practicable.

\begin{figure*}[bth]
    \centering
    \includegraphics[width=0.96\textwidth]{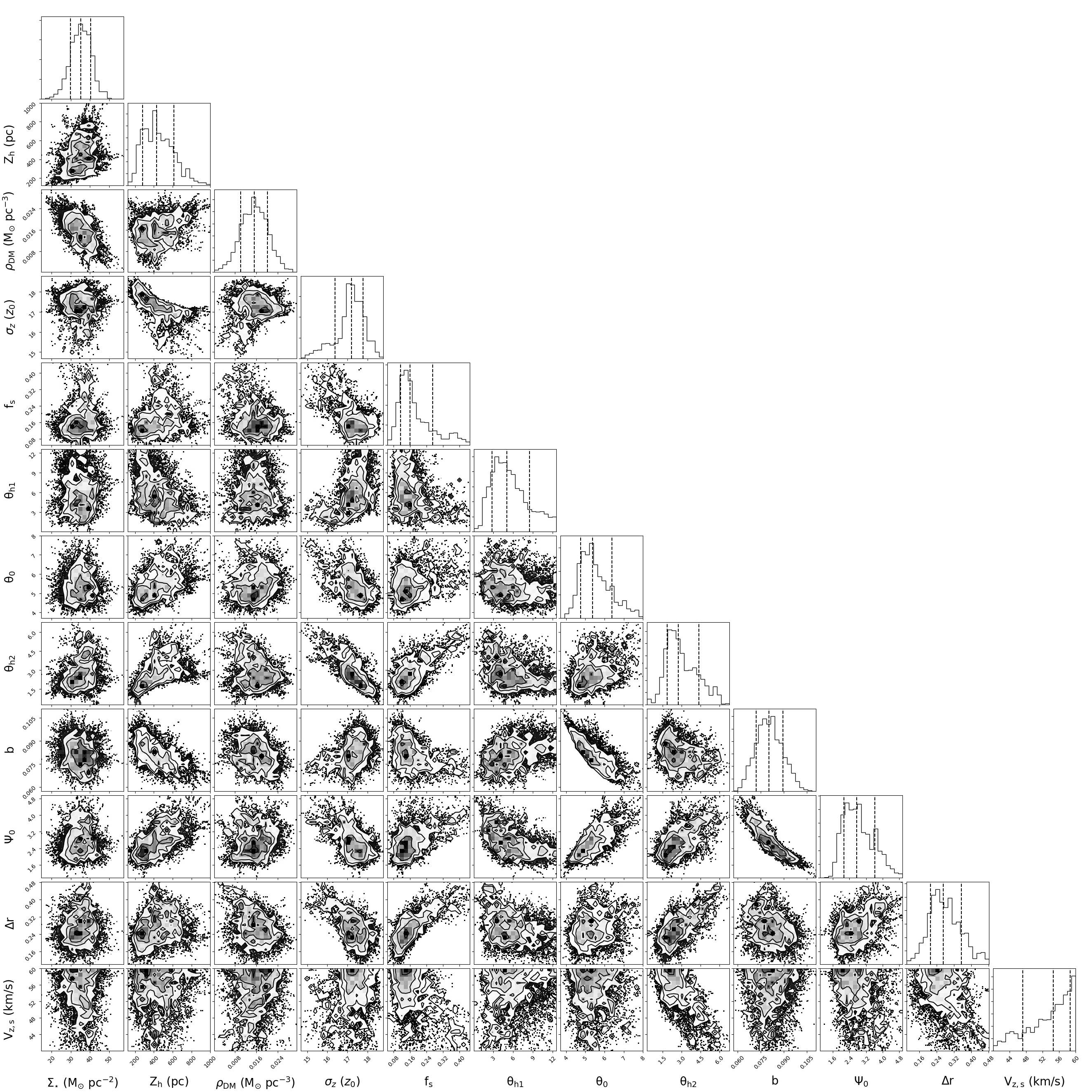}
    \caption{
    \label{fig:paras_dexp}
    Parameters' posterior distributions from MCMC (30,000 steps) for the `dexp' model.}
\end{figure*}






\newpage
\begin{thebibliography}{99}

\bibitem[\protect\citeauthoryear{Antoja et al.}{2018}]{Antoja2018} Antoja T., Helmi A., Romero-G{\'o}mez M., Katz D., Babusiaux C., Drimmel R., Evans D.~W., et al., 2018, Natur, 561, 360. doi:10.1038/s41586-018-0510-7
\bibitem[\protect\citeauthoryear{Bahcall}{1984}]{Bahcall1984} Bahcall J.~N., 1984, ApJ, 276, 169. doi:10.1086/161601
\bibitem[\protect\citeauthoryear{Banik, Widrow, \& Dodelson}{2017}]{Banik2017} Banik N., Widrow L.~M., Dodelson S., 2017, MNRAS, 464, 3775. doi:10.1093/mnras/stw2603
\bibitem[\protect\citeauthoryear{Bennett \& Bovy}{2019}]{Bennett2019} Bennett M., Bovy J., 2019, MNRAS, 482, 1417. doi:10.1093/mnras/sty2813
\bibitem[\protect\citeauthoryear{Bienaym{\'e} et al.}{2014}]{Bienayme2014} Bienaym{\'e} O., Famaey B., Siebert A., Freeman K.~C., Gibson B.~K., Gilmore G., Grebel E.~K., et al., 2014, A\&A, 571, A92. doi:10.1051/0004-6361/201424478
\bibitem[Binney \& Tremaine(2008)]{Binney2008} Binney, J. \& Tremaine, S.\ 2008, Galactic Dynamics: Second Edition, by James Binney and Scott Tremaine. ISBN 978-0-691-13026-2 (HB). Published by Princeton University Press, Princeton, NJ USA, 2008
\bibitem[Binney(2010)]{Binney2010} Binney, J.\ 2010, \mnras, 401, 2318. doi:10.1111/j.1365-2966.2009.15845.x
\bibitem[Binney \& McMillan(2011)]{Binney2011} Binney, J. \& McMillan, P.\ 2011, \mnras, 413, 1889. doi:10.1111/j.1365-2966.2011.18268.x
\bibitem[Binney \& Sch{\"o}nrich(2018)]{Binney2018} Binney, J. \& Sch{\"o}nrich, R.\ 2018, \mnras, 481, 1501. doi:10.1093/mnras/sty2378
\bibitem[Bland-Hawthorn et al.(2019)]{Bland2019} Bland-Hawthorn, J., Sharma, S., Tepper-Garcia, T., et al.\ 2019, \mnras, 486, 1167. doi:10.1093/mnras/stz217
\bibitem[Bland-Hawthorn \& Tepper-Garc{\'\i}a(2021)]{Bland2021} Bland-Hawthorn, J. \& Tepper-Garc{\'\i}a, T.\ 2021, \mnras, 504, 3168. doi:10.1093/mnras/stab704
\bibitem[\protect\citeauthoryear{Bovy \& Tremaine}{2012}]{Bovy2012} Bovy J., Tremaine S., 2012, ApJ, 756, 89. doi:10.1088/0004-637X/756/1/89
\bibitem[Bovy \& Rix(2013)]{Bovy2013} Bovy, J. \& Rix, H.-W.\ 2013, \apj, 779, 115. doi:10.1088/0004-637X/779/2/115
\bibitem[\protect\citeauthoryear{Carlin et al.}{2013}]{Carlin2013} Carlin J.~L., DeLaunay J., Newberg H.~J., Deng L., Gole D., Grabowski K., Jin G., et al., 2013, ApJL, 777, L5. doi:10.1088/2041-8205/777/1/L5
\bibitem[\protect\citeauthoryear{Carrillo et al.}{2018}]{Carrillo2018} Carrillo I., Minchev I., Kordopatis G., Steinmetz M., Binney J., Anders F., Bienaym{\'e} O., et al., 2018, MNRAS, 475, 2679. doi:10.1093/mnras/stx3342
\bibitem[Cautun et al.(2020)]{Cautun2020} Cautun, M., Ben{\'\i}tez-Llambay, A., Deason, A.~J., et al.\ 2020, \mnras, 494, 4291. doi:10.1093/mnras/staa1017
\bibitem[Chen et al.(1999)]{ChenB1999} Chen, B., Figueras, F., Torra, J., et al.\ 1999, \aap, 352, 459
\bibitem[Cheng et al.(2019)]{ChengXL2019} Cheng, X., Liu, C., Mao, S., et al.\ 2019, \apjl, 872, L1. doi:10.3847/2041-8213/ab020e
\bibitem[\protect\citeauthoryear{Chequers, Widrow, \& Darling}{2018}]{Chequers2018} Chequers M.~H., Widrow L.~M., Darling K., 2018, MNRAS, 480, 4244. doi:10.1093/mnras/sty2114
\bibitem[Cui et al.(2012)]{CuiX2012} Cui, X.-Q., Zhao, Y.-H., Chu, Y.-Q., et al.\ 2012, Research in Astronomy and Astrophysics, 12, 1197. doi:10.1088/1674-4527/12/9/003
\bibitem[Darling \& Widrow(2019)]{Darling2019} Darling, K. \& Widrow, L.~M.\ 2019, \mnras, 484, 1050. doi:10.1093/mnras/sty3508
\bibitem[\protect\citeauthoryear{Debattista}{2014}]{Debattista2014} Debattista V.~P., 2014, MNRAS, 443, L1. doi:10.1093/mnrasl/slu069
\bibitem[de Salas et al.(2019)]{Salas2019} de Salas, P.~F., Malhan, K., Freese, K., et al.\ 2019, \jcap, 2019, 037. doi:10.1088/1475-7516/2019/10/037
\bibitem[de Salas \& Widmark(2021)]{Salas2020} de Salas, P.~F. \& Widmark, A.\ 2021, Reports on Progress in Physics, 84, 104901. doi:10.1088/1361-6633/ac24e7
\bibitem[\protect\citeauthoryear{Faure, Siebert, \& Famaey}{2014}]{Faure2014} Faure C., Siebert A., Famaey B., 2014, MNRAS, 440, 2564. doi:10.1093/mnras/stu428
\bibitem[Flynn et al.(2006)]{Flynn2006} Flynn, C., Holmberg, J., Portinari, L., et al.\ 2006, \mnras, 372, 1149. doi:10.1111/j.1365-2966.2006.10911.x
\bibitem[Foreman-Mackey et al.(2013)]{Foreman2013} Foreman-Mackey, D., Hogg, D.~W., Lang, D., et al.\ 2013, \pasp, 125, 306. doi:10.1086/670067
\bibitem[\protect\citeauthoryear{Gaia Collaboration et al.}{2018a}]{Gaia2018a} Gaia Collaboration, Katz D., Antoja T., Romero-G{\'o}mez M., Drimmel R., Reyl{\'e} C., Seabroke G.~M., et al., 2018a, A\&A, 616, A11. doi:10.1051/0004-6361/201832865
\bibitem[\protect\citeauthoryear{Gaia Collaboration et al.}{2018b}]{Gaia2018b} Gaia Collaboration, Brown A.~G.~A., Vallenari A., Prusti T., de Bruijne J.~H.~J., Babusiaux C., Bailer-Jones C.~A.~L., et al., 2018b, A\&A, 616, A1. doi:10.1051/0004-6361/201833051
\bibitem[Gao et al.(2014)]{GaoS2014} Gao, S., Liu, C., Zhang, X., et al.\ 2014, \apjl, 788, L37. doi:10.1088/2041-8205/788/2/L37
\bibitem[\protect\citeauthoryear{Garbari et al.}{2012}]{Garbari2012} Garbari S., Liu C., Read J.~I., Lake G., 2012, MNRAS, 425, 1445. doi:10.1111/j.1365-2966.2012.21608.x
\bibitem[\protect\citeauthoryear{G{\'o}mez et al.}{2013}]{Gomez2013} G{\'o}mez F.~A., Minchev I., O'Shea B.~W., Beers T.~C., Bullock J.~S., Purcell C.~W., 2013, MNRAS, 429, 159. doi:10.1093/mnras/sts327
\bibitem[\protect\citeauthoryear{Guo et al.}{2020}]{Guo2020} Guo R., Liu C., Mao S., Xue X.-X., Long R.~J., Zhang L., 2020, MNRAS, 495, 4828. doi:10.1093/mnras/staa1483
\bibitem[Hagen \& Helmi(2018)]{Hagen2018} Hagen, J.~H.~J. \& Helmi, A.\ 2018, \aap, 615, A99. doi:10.1051/0004-6361/201832903
\bibitem[\protect\citeauthoryear{Haines et al.}{2019}]{Haines2019} Haines T., D'Onghia E., Famaey B., Laporte C., Hernquist L., 2019, ApJL, 879, L15. doi:10.3847/2041-8213/ab25f3
\bibitem[\protect\citeauthoryear{Holmberg \& Flynn}{2000}]{Holmberg2000} Holmberg J., Flynn C., 2000, MNRAS, 313, 209. doi:10.1046/j.1365-8711.2000.02905.x
\bibitem[\protect\citeauthoryear{Holmberg \& Flynn}{2004}]{Holmberg2004} Holmberg J., Flynn C., 2004, MNRAS, 352, 440. doi:10.1111/j.1365-2966.2004.07931.x
\bibitem[Huang et al.(2016)]{Huang2016} Huang, Y., Liu, X.-W., Yuan, H.-B., et al.\ 2016, \mnras, 463, 2623. doi:10.1093/mnras/stw2096
\bibitem[Irrgang et al.(2013)]{Irrgang2013} Irrgang, A., Wilcox, B., Tucker, E., et al.\ 2013, \aap, 549, A137. doi:10.1051/0004-6361/201220540
\bibitem[Khoperskov et al.(2019)]{Khoperskov2019} Khoperskov, S., Di Matteo, P., Gerhard, O., et al.\ 2019, \aap, 622, L6. doi:10.1051/0004-6361/201834707
\bibitem[\protect\citeauthoryear{Kuijken \& Gilmore}{1989a}]{Kuijken1989a} Kuijken K., Gilmore G., 1989a, MNRAS, 239, 571. doi:10.1093/mnras/239.2.571
\bibitem[\protect\citeauthoryear{Kuijken \& Gilmore}{1989b}]{Kuijken1989b} Kuijken K., Gilmore G., 1989b, MNRAS, 239, 605. doi:10.1093/mnras/239.2.605
\bibitem[\protect\citeauthoryear{Laporte et al.}{2018}]{Laporte2018} Laporte C.~F.~P., Johnston K.~V., G{\'o}mez F.~A., Garavito-Camargo N., Besla G., 2018, MNRAS, 481, 286. doi:10.1093/mnras/sty1574
\bibitem[Laporte et al.(2019)]{Laporte2019} Laporte, C.~F.~P., Minchev, I., Johnston, K.~V., et al.\ 2019, \mnras, 485, 3134. doi:10.1093/mnras/stz583
\bibitem[Li \& Widrow(2021)]{LiH2021} Li, H. \& Widrow, L.~M.\ 2021, \mnras, 503, 1586. doi:10.1093/mnras/stab574
\bibitem[Li \& Shen(2020)]{LiZY2020} Li, Z.-Y. \& Shen, J.\ 2020, \apj, 890, 85. doi:10.3847/1538-4357/ab6b21
\bibitem[Li(2021)]{LiZY2021} Li, Z.-Y.\ 2021, \apj, 911, 107. doi:10.3847/1538-4357/abea17
\bibitem[\protect\citeauthoryear{Monari, Famaey, \& Siebert}{2015}]{Monari2015} Monari G., Famaey B., Siebert A., 2015, MNRAS, 452, 747. doi:10.1093/mnras/stv1206
\bibitem[\protect\citeauthoryear{Monari et al.}{2016}]{Monari2016} Monari G., Famaey B., Siebert A., Grand R.~J.~J., Kawata D., Boily C., 2016, MNRAS, 461, 3835. doi:10.1093/mnras/stw1564
\bibitem[\protect\citeauthoryear{Oort}{1932}]{Oort1932} Oort J.~H., 1932, BAN, 6, 249
\bibitem[Purcell et al.(2011)]{Purcell2011} Purcell, C.~W., Bullock, J.~S., Tollerud, E.~J., et al.\ 2011, \nat, 477, 301. doi:10.1038/nature10417
\bibitem[\protect\citeauthoryear{Read}{2014}]{Read2014} Read J.~I., 2014, JPhG, 41, 063101. doi:10.1088/0954-3899/41/6/06310
\bibitem[Reid et al.(2014)]{Reid2014b} Reid, M.~J., Menten, K.~M., Brunthaler, A., et al.\ 2014, \apj, 783, 130. doi:10.1088/0004-637X/783/2/130
\bibitem[\protect\citeauthoryear{Salomon et al.}{2020}]{Salomon2020} Salomon J.-B., Bienaym{\'e} O., Reyl{\'e} C., Robin A.~C., Famaey B., 2020, A\&A, 643, A75. doi:10.1051/0004-6361/202038535
\bibitem[\protect\citeauthoryear{Sivertsson et al.}{2018}]{Sivertsson2018} Sivertsson S., Silverwood H., Read J.~I., Bertone G., Steger P., 2018, MNRAS, 478, 1677. doi:10.1093/mnras/sty977
\bibitem[Skrutskie et al.(2006)]{Skrutskie2006} Skrutskie, M.~F., Cutri, R.~M., Stiening, R., et al.\ 2006, \aj, 131, 1163. doi:10.1086/498708
\bibitem[Sun et al.(2015)]{SunNC2015} Sun, N.-C., Liu, X.-W., Huang, Y., et al.\ 2015, Research in Astronomy and Astrophysics, 15, 1342. doi:10.1088/1674-4527/15/8/017
\bibitem[Tian et al.(2015)]{TianH2015} Tian, H.-J., Liu, C., Carlin, J.~L., et al.\ 2015, \apj, 809, 145. doi:10.1088/0004-637X/809/2/145
\bibitem[Tian et al.(2018)]{Tian2018} Tian, H.-J., Liu, C., Wu, Y., et al.\ 2018, \apjl, 865, L19. doi:10.3847/2041-8213/aae1f3
\bibitem[Wang et al.(2019)]{WangC2019} Wang, C., Huang, Y., Yuan, H.-B., et al.\ 2019, \apjl, 877, L7. doi:10.3847/2041-8213/ab1fdd
\bibitem[\protect\citeauthoryear{Widmark}{2019}]{Widmark2019} Widmark A., 2019, A\&A, 623, A30. doi:10.1051/0004-6361/201834718
\bibitem[Widmark et al.(2021a)]{Widmark2021a} Widmark, A., Laporte, C., \& de Salas, P.~F.\ 2021, arXiv:2102.08955
\bibitem[Widmark et al.(2021b)]{Widmark2021b} Widmark, A., Laporte, C., de Salas, P.~F., et al.\ 2021, arXiv:2105.14030
\bibitem[\protect\citeauthoryear{Widrow et al.}{2012}]{Widrow2012} Widrow L.~M., Gardner S., Yanny B., Dodelson S., Chen H.-Y., 2012, ApJL, 750, L41. doi:10.1088/2041-8205/750/2/L41
\bibitem[\protect\citeauthoryear{Widrow et al.}{2014}]{Widrow2014} Widrow L.~M., Barber J., Chequers M.~H., Cheng E., 2014, MNRAS, 440, 1971. doi:10.1093/mnras/stu396
\bibitem[\protect\citeauthoryear{Williams et al.}{2013}]{Williams2013} Williams M.~E.~K., Steinmetz M., Binney J., Siebert A., Enke H., Famaey B., Minchev I., et al., 2013, MNRAS, 436, 101. doi:10.1093/mnras/stt1522
\bibitem[\protect\citeauthoryear{Xia et al.}{2016}]{Xia2016} Xia Q., Liu C., Mao S., Song Y., Zhang L., Long R.~J., Zhang Y., et al., 2016, MNRAS, 458, 3839. doi:10.1093/mnras/stw565
\bibitem[Xu et al.(2020)]{XuY2020} Xu, Y., Liu, C., Tian, H., et al.\ 2020, \apj, 905, 6. doi:10.3847/1538-4357/abc2cb
\bibitem[\protect\citeauthoryear{Yanny \& Gardner}{2013}]{Yanny2013} Yanny B., Gardner S., 2013, ApJ, 777, 91. doi:10.1088/0004-637X/777/2/91
\bibitem[\protect\citeauthoryear{Zhang et al.}{2013}]{Zhang2013} Zhang L., Rix H.-W., van de Ven G., Bovy J., Liu C., Zhao G., 2013, ApJ, 772, 108. doi:10.1088/0004-637X/772/2/108
\bibitem[Zhao et al.(2012)]{ZhaoG2012} Zhao, G., Zhao, Y.-H., Chu, Y.-Q., et al.\ 2012, Research in Astronomy and Astrophysics, 12, 723. doi:10.1088/1674-4527/12/7/002

\end{thebibliography}
\end{document}